\documentclass[preprint,showpacs,preprintnumbers,amsmath,amssymb,superscriptaddress,12pt,graphicx]{revtex4}

\usepackage{graphicx}
\usepackage{dcolumn}
\usepackage{bm}
\usepackage{amssymb}

\renewcommand{\L}{{\cal L}}

\begin{document}

\title{Detecting relic gravitational waves in the CMB: Comparison of different
methods}

\author{W.~Zhao}
\small\email{Wen.Zhao@astro.cf.ac.uk} \affiliation{School of Physics
and Astronomy, Cardiff University, Cardiff, CF24 3AA, United
Kingdom} \affiliation{Wales Institute of Mathematical and
Computational Sciences, Swansea, SA2 8PP, United Kingdom} \affiliation{Department of
Physics, Zhejiang University of Technology, Hangzhou, 310014,
P. R. China}



\begin{abstract}
{\small In this paper, we discuss the constraint on the relic
gravitational waves by both temperature and polarization
anisotropies power spectra of cosmic microwave background
radiation. Taking into account the instrumental noises of Planck
satellite, we calculate the signal-to-noise ratio $S/N$ by the
simulation and the analytic approximation methods. We find that,
comparing with the $BB$ channel, the value of $S/N$ is much
improved in the case where all the power spectra, $TT$, $TE$, $EE$
and $BB$, are considered. If the noise power spectra of Planck
satellite increase for some reasons, the value of $S/N$ in $BB$
channel is much reduced. However, in the latter case where all the
power spectra of cosmic microwave background radiation are
considered, the value of $S/N$ is less influenced. We also find
that the free parameters $A_s$, $n_s$ and $n_t$ have little
influence on the value of $S/N$ in both cases.}

\end{abstract}


\maketitle


\section{Introduction \label{introduction}}

A stochastic background of the relic gravitational waves (RGWs),
generated during the very early stage of the Universe \cite{a1},
is a necessity dictated by general relativity and quantum
mechanics \cite{grishchuk0707}. The RGWs have a wide range
spreading spectra \cite{grishchuk1,zhang1,others}, and their
detection plays a double role in relativity and cosmology.

One of the important methods for the detection of RGWs is by the
cosmic microwave background (CMB) power spectra, including the
temperature anisotropies ($TT$) power spectrum,  the polarization
($EE$ and $BB$) power spectra, and the cross-correlation ($TE$)
power spectrum \cite{a8,grishchuk-cmb,a11,a12,a13}. By observing
the lower-order CMB multipoles, one can detect the signal of RGWs
at the very low frequencies ($10^{-17}\sim10^{-15}$ Hz). One has
to note, besides the gravitational waves of quantum-mechanical
origin \cite{grishchuk0707}, the classical gravitational waves
were generated at later stages of cosmological evolution
\cite{othersource}. However, their wavelengths are much shorter
than the present Hubble radius and therefore they do not affect
the lower-order CMB multipoles.

As well known, the CMB has certain degree of polarization
generated via Thompson scattering during the decoupling in the early
Universe \cite{a5}.  In particular, if the RGWs are present at the
photon decoupling in the Universe, the magnetic type of
polarization ($B$-polarization) will be produced
\cite{a8,a11,a12}. This would be a characteristic feature of RGWs
on very large scale, since the density perturbations will not
generate this polarization. So a natural way for the detection of
RGWs is by observing the signal of $B$-polarization of CMB. This
is the so-called ``{\it BB}" method. However, the amplitude of the
$B$-polarization is expected to be very small. In addition, the
$B$-polarization is prone to degradation by various systematic
effects on a wide range of scales
\cite{wuran1,wuran2,wuran3,wuran4}. The current 5-year Wilkinson
Microwave Anisotropy Probe (WMAP5) observation only gives an upper
limit $\ell(\ell+1)C_{\ell=2-6}^{BB}/2\pi<0.15\mu$K$^2$
($95\%$C.L.)\cite{5map}. The forthcoming projects, such as the
Planck \cite{Planck}, Clover \cite{clover}, Spider \cite{spider},
QUITE \cite{quiet}, are expected to be much more sensitive for the
detection of the CMB $B$-polarization.

Due to the disadvantage of ``{\it BB}" method, it is necessary to look
for the new method for the detection of RGWs in the CMB. In the
previous work \cite{a12}, the authors found that, in the large
scale $(\ell<50)$, the RGWs generate the negative $TE$ spectrum.
However, if the $TE$ spectrum is generated by density
perturbations, it should be positive. This suggests that the
signal of RGWs can also be detected by the CMB $TE$ spectrum.
Comparing with the $B$-polarization, the amplitude of the $TE$
spectrum is nearly two order larger.  In the works
\cite{polnarev,ours}, the authors have developed several ways to
detect the signal of RGWs directly from the CMB $TE$ spectrum.
These are the so-called ``{\it TE}" method.

In the work \cite{ours} we found the WMAP5 $TE$ data contains a
hint of the presence of RGWs contribution. In terms of quadrupole
ratio $R$, the best-fit model produced $R=0.24$, which corresponds
to the tensor-to-scalar ratio $r\simeq 0.48$. Because of large
residual noise, the uncertainty of this determination is still
large. We also found, if considering the Planck instrumental
noises, ``{\it TE}" method can detect the signal of RGWs at 2$\sigma$
level when $r>0.3$. If considering the ideal case with full sky
and no noise, ``{\it TE}" method can detect the RGWs at 2$\sigma$ level
when $r>0.1$. By comparing the detection abilities of ``{\it TE}" and
``{\it BB}" methods, we found that, taking into account of the
instrumental noises of Planck satellite, ``{\it BB}" method is more
sensitive when $r$ is small. However, if the noise power spectra
or the amplitudes of RGWs increase, the sensitivity of ``{\it TE}"
method becomes better than that of ``{\it BB}" method.

In this paper, we shall extend the ``{\it TE}" method in our previous
work \cite{ours}. By calculating the values of $S/N$, we shall
compare the detection abilities for the RGWs in the following four
cases. The first case (``{\it B}" case) is the so-called ``{\it BB}" method,
where only the $B$-polarization spectrum is considered. In the
second case, we include not only the CMB $TE$ spectrum, but also
the $TT$ spectrum. We call it as the ``{\it CT}" case (``C" standing for
the cross-correlation power spectrum and ``T" standing for the
temperature anisotropies power spectrum).  We shall expect that,
taking into account the contribution of $TT$ spectrum, the
detection ability will be much improved. In the third case,
besides $TE$ and $TT$ spectra, we also include the $EE$ spectrum.
We call it as the ``{\it CTE}" case. The fourth one is the ``{\it CTEB}" case,
where the contributions of $TE$, $TT$, $EE$ and $BB$ power spectra
are all considered. The detection ability in this case is expected
to be much more sensitive than the other three cases.

The organization of this paper is as follows. In Section II, the
primordial power spectra of RGWs and density perturbations, the
CMB power spectra and the corresponding estimators are introduced.
In this section, the probability density functions (pdfs) for the
estimators are also discussed. In Section III, we introduce four
(``{\it B}", `CT", ``{\it CTE}" and ``{\it CTEB}") cases  for the detection of RGWs
in the CMB. The likelihood functions are also given this section.
In Section IV, by constructing the likelihood functions based on
the simulated data, we shall investigate the values of the
signal-to-noise ratio $S/N$ in these four cases. This quantity
describes the detection abilities of RGWs in the different cases.
We firstly introduce the simulation method. In the simulation,
when constructing the likelihood functions, we only consider one
free cosmic parameter, the tensor-to-scalar ratio $r$. We find
when $r>0.06$ ($r>0.16$, $r>0.13$, $r>0.05$), the signal of RGWs
can be detected in ``{\it B}" (``{\it CT}", ``{\it CTE}", ``{\it CTEB}") case at $2\sigma$
level. In Section V, we discuss the analytic approximation of the
likelihood functions. By the analytic approximation formulae, we
obtain a simple analytic form of $S/N$, which clearly shows the
dependence of $S/N$ on the amplitude of RGWs and the noises. By
analyzing the analytic form of $S/N$, we find that, in ``{\it CT}" and
``{\it CTE}" cases, the main contributions come from the data in the
intermedial scale $10<\ell<70$. However, in ``{\it B}" case, the main
contribution comes from the data in the very large scale
$\ell<10$.  In Section VI, by the simulation method, we find that,
the free parameters (the tensor spectral index $n_t$, the
amplitude of scalar spectrum $A_s$ and the scalar spectral index
$n_s$) have little influence on the determination of RGWs. Section
VII is the conclusion that summarizes the main results of this
paper.


\section{Gravitational field perturbations, CMB power spectra and their estimators \label{section2}}


\subsection{Primordial power spectra of the gravitational field perturbations\label{section2.0}}

The CMB temperature and polarization anisotropies power spectra
are determined by the primordial power spectra of density
perturbations (scalar perturbations) and RGWs (tensor
perturbations), and the time evolution of these perturbations
during and after the epoch of recombination. Before proceeding
with the CMB power spectra, it is necessary to introduce the
primordial perturbation spectra, which are usually assumed to be
power-law. This form is a generic prediction of a wide range of
scenarios of the early universe \cite{a1,a4,grishchuk1}. In
general there might be deviations from a power-law, parametrized
in terms of the running of the spectral index (see for example
\cite{liddle}), but we shall not consider this probability in the
current paper. Thus the primordial power spectra of the
perturbation fields have the forms
\begin{eqnarray}\label{PsPt}
\begin{array}{c}
P_{s}(k) = A_{s}(k_0)
\left(\frac{k}{k_0}\right)^{n_s-1}, ~~~ P_t (k)= A_t(k_0)
\left(\frac{k}{k_0}\right)^{n_t},
\end{array}
\end{eqnarray}
where $n_s$ and $n_t$ are the scalar and tensor spectral indices,
respectively. $k_0$ is pivot wavenumber, which can be arbitrarily
chosen.  In the WMAP paper \cite{peiris}, the pivot wavenumber
$k_0=0.002$Mpc$^{-1}$ is used, which is close to the observable
horizon. The scale $k_0=0.05$Mpc$^{-1}$  is also commonly used,
being the default scale of CAMB package \cite{cosmomc}.  A number
of authors have suggested other pivot wavenumber $k_0$ for
different reasons \cite{pivot1}. In Eq. (\ref{PsPt}), $A_s(k_0)$
and $A_t(k_0)$ are the amplitudes of the primordial scalar and
tensor spectra respectively, at the pivot scale $k_0$.

We can re-parameterize the tensor power spectrum amplitude
$A_t(k_0)$ by the ``tensor-to-scalar ratio" $r$, which is defined
by
\begin{eqnarray}\label{define_r}
r(k_0)\equiv \frac{A_t(k_0)}{A_s(k_0)}.
\end{eqnarray}
In addition, the ratio of tensor quadrupole to scalar quadrupole
$R$ is also quoted when referring to the tensor-to-scalar ratio
(see for instant \cite{knox,ours}). The relation between $R$ and
$r$ is somewhat cosmology-dependence, especially on the dark
energy density $\Omega_{\Lambda}$. The conversion is $R\simeq
\frac{0.84-0.025\Omega_{\Lambda}-0.084\Omega_{\Lambda}^2}{1.04-0.82\Omega_{\Lambda}+2\Omega_{\Lambda}^2}r$
\cite{R_r}. For the cosmological models with
$\Omega_{\Lambda}=0.76$, these two definitions are simply related
by $r\simeq 2R$. In the previous work \cite{ours}, we have adopted
$R$. However, in this paper we shall use $r$, the default quantity
used in the CAMB package \cite{camb}.

Using Eqs. (\ref{PsPt}) and (\ref{define_r}), one can evaluate $r$
at a different wavenumber $k_1$,
\begin{eqnarray}\label{k1k0}
r(k_1)=r(k_0)\left(\frac{k_1}{k_0}\right)^{n_t-n_s+1}.
\end{eqnarray}
In the following discussion, we shall discuss the constraint on
the parameter $r$ by the simulated data. From the relation
(\ref{k1k0}) we find that, if the spectral indices $n_t$ and $n_s$
are fixed as $n_t=0$ and $n_s=1$ in the likelihood analysis (the
case in Sections \ref{s3} and \ref{s4}), we have $r(k_1)\equiv
r(k_0)$, the tensor-to-scalar ratio $r$ has the exactly same value
at all pivot wavenumbers. So we do not need to differentiate the
tensor-to-scalar ratio at the different pivot scales. However, if
the spectral indices are free parameters in the likelihood
analysis (the case in Section \ref{s5}), comparing with $r(k_0)$,
the constraint of $r(k_1)$ is also influenced by the spectral
indices. So in this case, we should differentiate the
tensor-to-scalar ratio at the different pivot wavenumbers. This
effect will be clearly shown in Section \ref{s5}.

We should mention that, in the specific early universe models, the
parameters $A_s$, $n_s$, $r$ and $n_t$ are always not separate
\cite{relation}. However, in this paper we shall avoid any
specific model, and consider the parameters $A_s$, $n_s$, $r$ and
$n_t$ as the independent parameters in the data analysis.


\subsection{CMB power spectra\label{section2.1}}

The CMB radiation field is usually characterized by four Stokes
parameters ($I,Q,U,V$). $I$ is the total intensity of radiation,
$Q$ and $U$ describe the magnitude and direction of linear
polarization, and $V$ is the circular polarization. From these
Stokes parameters, we can construct four invariant quantities
($I,V,E,B$), which can be expanded over ordinary spherical
harmonics (see \cite{a12} for details). The set of multipole
coefficients
$(a_{{\ell}m}^T,a_{{\ell}m}^V,a_{{\ell}m}^E,a_{{\ell}m}^B)$
completely characterize the intensity and polarization of the
radiation field. Since Thompson scattering of initial unpolarized
light cannot generate circular polarization, we shall not consider
the $V$ stokes parameter in the following discussion.

In general, the output of the CMB experiment $a_{\ell m}^X$
($X=T,E,B$), consists of two contributions, the signal convolved
with the beam window function $a_{\ell m}^X(s)W_{\ell}$ and the
noise $a_{\ell m}^X(n)$, i.e.
\begin{eqnarray}\label{s+n}
a_{\ell m}^X=a_{\ell m}^X (s)W_{\ell} +a_{\ell m}^X(n).
\end{eqnarray}
We shall use the notations $(s)$ and $(n)$ to denote the signal
and noise. These two contributions are uncorrelated to each other.

Assuming the primordial perturbation fields (including the scalar
perturbations and tensor perturbations) are Gaussian fields, which
induces that the signal term $a_{\ell m}^X(s)$ has the covariance
\cite{ours}
\begin{eqnarray}\label{powerspectra}
\frac{1}{2}\langle a_{\ell m}^{X}(s)a_{\ell' m'}^{X'*}(s)+a_{\ell
m}^{X*}(s)a_{\ell'
m'}^{X'}(s)\rangle=C_{\ell}^{XX'}\delta_{\ell\ell'}\delta_{mm'},
\end{eqnarray}
where $C_{\ell}^{XX'}$ is known as the CMB power spectra, which
depends on the cosmological inputs. When $X=X'$, $C_{\ell}^{XX'}$
is the auto-correlation power spectra, and when $X\neq X'$,
$C_{\ell}^{XX'}$ is the cross-correlation power spectra. In
absence of any parity-violating processes, the only survived
cross-correlation power spectra is $C_{\ell}^{TE}$ \cite{lue}. So
the temperature and polarization anisotropies can be described
completely by four power spectra: $C_{\ell}^{TT}$,
$C_{\ell}^{EE}$, $C_{\ell}^{BB}$ and $C_{\ell}^{TE}$.

The noise terms $a_{\ell m}^X(n)$ and the window function
$W_{\ell}$ depend on the experiment. We assume the noise is a
spatially uniform Gaussian white noise. For an experiment with
some known beam width and sensitivity, the noise power spectra and
window function can be approximated as
\begin{subequations}
\begin{eqnarray}\label{noises}
N_{\ell}^{XX'}=\frac{1}{2}\langle a_{\ell m}^{X}(n) a_{\ell'
m'}^{X'*}(n)+a_{\ell m}^{X*}(n) a_{\ell'
m'}^{X'}(n)\rangle=(\theta_{\rm
FWHM}\sigma_X)^2\delta_{XX'}\delta_{\ell\ell'}\delta_{mm'},
\end{eqnarray}
\begin{eqnarray}\label{window}
W_{\ell}=\exp\left[-\frac{\ell(\ell+1)}{2}\frac{\theta^2_{\rm
FWHM}}{8\ln2}\right],
\end{eqnarray}
where $\theta_{\rm FWHM}$ is the full width at half maximum of the
Gaussian beam, and $\sigma_X$ is the root mean square of the
instrumental noise. Non-diagonal noise terms (i.e., $X\neq X'$)
are expected to vanish since the noises contributions from
different maps are uncorrelated. The assumption of a spatially
uniform Gaussian noises spectrum ensures that the noise term is
diagonal in the $\ell$ basis. In this paper, we shall consider the
Planck instrumental noises. There are several frequency channels
for the detection of CMB in Planck satellite \cite{Planck}. In
this paper, in order to estimate the Planck noises, we only adopt
the frequency channel at 143GHz, which has the low foreground
levels and the lowest noises power spectra. In this channel, we
have \cite{Planck},
\begin{eqnarray}\label{Plancknoises}
\theta_{\rm FWHM}=7.1',~~\sigma_T=6.0{\rm \mu K},
~~\sigma_E=\sigma_B=11.5{\rm \mu K}.
\end{eqnarray}
Inserting these into Eq. (\ref{noises}), we obtain the noise power
spectra
\begin{eqnarray}\label{noises2}
N_{\ell}^{TT}=1.53\times 10^{-4}{\rm \mu
K}^2,~~N_{\ell}^{EE}=N_{\ell}^{BB}=5.58\times 10^{-4}{\rm \mu
K}^2,~~N_{\ell}^{TE}=0.
\end{eqnarray}
\end{subequations}

Considering Eqs. (\ref{s+n}), (\ref{powerspectra}) and
(\ref{noises}), we obtain the covariances of the terms $a_{\ell
m}^{X}$, which are
\begin{eqnarray}\label{output}
\frac{1}{2}\langle a_{\ell m}^X a_{\ell' m'}^{X'*}+a_{\ell m}^{X*}
a_{\ell' m'}^{X'}\rangle=(C_{\ell}^{XX'}
W_{\ell}^2+N_{\ell}^{XX'})\delta_{\ell\ell'}\delta_{mm'}.
\end{eqnarray}


\subsection{Estimators of the CMB power spectra}

In Section \ref{section2.1}, we have introduced the CMB power
spectra, which are defined as ensemble averages over all possible
realization of the CMB field. However, in CMB observations, we
only have access to one single realization of this ensemble. In
order to obtain information on the power spectra from a single
realization, it is desirable to introduce the estimators of the
power spectra, which are observable quantities.

In the full sky case, and taking into account the noises, the best
unbiased estimators $D_{\ell}^{XX'}$ for the CMB power spectra
$C_{\ell}^{XX'}$ are defined by \cite{grishchuk,ours}
   \begin{eqnarray}\label{define_estimator}
 D_{\ell}^{XX'}=\left(\frac{1}{n}\sum_{m=-\ell}^{\ell}(a_{\ell m}^{X}a_{\ell
m}^{X'*}+a_{\ell m}^{X*}a_{\ell
m}^{X'})-N_{\ell}^{XX'}\right)W_{\ell}^{-2}~,
 \end{eqnarray}
where $n$ is the number of the degree of freedom for a fix multipole
$\ell$. In the full sky case, we have $n=(2\ell+1)$. The expectation values and the standard deviations of
these estimators are \cite{ours}
\begin{eqnarray}\label{mean--variances}
\begin{array}{c}
 \langle D_{\ell}^{XX'}\rangle=C_{\ell}^{XX'}, \\ \Delta D_{\ell}^{XX'}=

\sqrt{\frac{(C_{\ell}^{XX}+N_{\ell}^{XX}W_{\ell}^{-2})(C_{\ell}^{X'X'}+N_{\ell}^{X
'X'}W_{\ell}^{-2})+(C_{\ell}^{XX'}+N_{\ell}^{XX'}W_{\ell}^{-2})^2}{n}}.
 \end{array}
 \end{eqnarray}

It is necessary to investigate the pdfs for $D_{\ell}^{XX'}$,
which have been derived in \cite{ours}, based on the assumption:
the primordial perturbation fields and noise fields are
independent Gaussian fields. In this subsection, we shall briefly
introduce the results as follows (the similar results are also
obtained in the Refs. \cite{wishart2,pdf1}).

The pdf of the auto-correlation estimator $D_{\ell}^{XX}$ is known
as the $\chi^2$ distribution, which is
 \begin{eqnarray}\label{chi2}
f(D_{\ell}^{XX})=\frac{(nW^2_{\ell})V^{(n-2)/2}e^{-V/2}}{2^{n/2}\Gamma(n/2)(C_{\ell}^{XX}W_{\ell}^2+N_{\ell}^{XX})},
 \end{eqnarray}
where $n = (2\ell+1)$ is the degree of freedom for the multipole
$\ell$ in the full sky case. The quantity $V$ is defined by
$V\equiv
n(D_{\ell}^{XX}W_{\ell}^2+N_{\ell}^{XX})/(C_{\ell}^{XX}W_{\ell}^2+N_{\ell}^{XX})$.

The joint pdf for the estimators $D_{\ell}^{TE}$, $D_{\ell}^{TT}$
and $D_{\ell}^{EE}$ is the following Wishart distribution
 \begin{eqnarray}\label{wishart}
 \begin{array}{c}
f(D_{\ell}^{TE},D_{\ell}^{TT},D_{\ell}^{EE})=\left\{\frac{1}{4(1-{\rho_{\ell}^2})(
{\sigma_\ell^T}{\sigma_\ell^E})^2}\right\}^{n/2} \frac{
(nW_{\ell}^2)^3({x}{y}-{z}^2)^{(n-3)/2}}{\pi^{1/2}\Gamma(n/2)\Gamma((n-1)/2)} \\
\times
\exp\left\{-\frac{1}{2(1-{\rho_\ell^2})}\left(\frac{{x}}{(\sigma_\ell^T)^2}+\frac{
{y}}{(\sigma_\ell^E)^2}-\frac{2{\rho_l}
{z}}{{\sigma_\ell^T}{\sigma_\ell^E}}\right)\right\},
 \end{array}
 \end{eqnarray}
where the quantities $x,y,z$ are defined by: ${x}\equiv
n(D_\ell^{TT}W_{\ell}^2+N_{\ell}^{TT})$, ${y}\equiv
n(D_\ell^{EE}W_{\ell}^2+N_{\ell}^{EE})$, ${z}\equiv
nD_\ell^{TE}W_{\ell}^2$. ${\sigma_\ell^T}= \sqrt{
C_\ell^{TT}W_{\ell}^2+N_\ell^{TT}}$, $\sigma_\ell^E=\sqrt{
C_\ell^{EE}W_{\ell}^2+N_{\ell}^{EE}}$ are the standard deviations
of the multipole coefficients $a_{\ell m}^{T}$ and $a_{\ell
m}^{E}$, respectively. $\rho_{\ell}$ is the correlation
coefficient of $a_{\ell m}^{T}$ and $a_{\ell m}^{E}$, which can be
written as,
 \begin{eqnarray}\label{rho}
{\rho_\ell}=\frac{C_\ell^{TE}}{\sqrt{(C_\ell^{TT}+N_{\ell}^{TT}W_{\ell}^{-2})(C_\ell^{EE}+N_{\ell}^{EE}W_{\ell}^{-2})}}.
 \end{eqnarray}

From the Wishart distribution (\ref{wishart}), we can derive the
joint pdf of the estimators $D_{\ell}^{TE}$ and $D_{\ell}^{TT}$ by
integrating the variable  $D_{\ell}^{EE}$, the finial result is
 \begin{eqnarray} \label{pdf_CT}
 \begin{array}{c}
f(D_{\ell}^{TE},D_{\ell}^{TT})= (nW_{\ell}^2)^2{x}^{\frac{n-3}{2}}
\left\{2^{1+n}\pi\Gamma^2(\frac{n}{2})(1-\rho_{\ell}^2)(\sigma_{\ell}^T)^{2n}(\sigma_{\ell}^E)^2\right\}^{-\frac{1}{2}}
\\ \times\exp\left\{\frac{1}{1-\rho^2_{\ell}}\left(\frac{{\rho_{\ell}}
{z}}{{\sigma_\ell^T}{\sigma_\ell^E}}-\frac{{z}^2}{2x{(\sigma_\ell^E)^2}
}-\frac{{x}}{2{(\sigma_\ell^T)^2}}\right)\right\}.
\end{array}
 \end{eqnarray}

We can also obtain the joint pdf for all the four estimators:
$D_{\ell}^{TE}$, $D_{\ell}^{TT}$, $D_{\ell}^{EE}$,
$D_{\ell}^{BB}$. Since $B$-polarization estimator $D_{\ell}^{BB}$
is independent of the estimators $D_{\ell}^{TE}$, $D_{\ell}^{TT}$
and $D_{\ell}^{EE}$, the total joint pdf is the product of the
Wishart distribution
$f(D_{\ell}^{TE},D_{\ell}^{TT},D_{\ell}^{EE})$ in (\ref{wishart})
and the $\chi^2$ distribution $f(D_{\ell}^{BB})$ in (\ref{chi2})
with $XX=BB$, i.e.
 \begin{eqnarray}\label{joint}
 f(D_{\ell}^{TE},D_{\ell}^{TT},D_{\ell}^{EE},D_{\ell}^{BB})=
 f(D_{\ell}^{TE},D_{\ell}^{TT},D_{\ell}^{EE})f(D_{\ell}^{BB}).
 \end{eqnarray}

We should notice that, the above results are all based on the
assumption of full sky coverage. However, real experiments can
only see a fraction of sky. Even for satellite experiments, a map
cut must be performed in order to eliminate point sources and
galactic plane foreground contaminations. As a result, different
multipole moments ${a_{\ell m}^{X}}$ become correlated with each
other \cite{cut1,wuran2}. The exact pdfs of the estimators in this
case takes a rather complicated form, depending on the shape of
remaining observed portion of sky \cite{wishart2}. However, for
experiments probing almost the full sky (e.g. COBE, WMAP, or
Planck), correlations are expected only between neighboring
multipoles. In order to simplify the problem, one can take
${a_{\ell m}^{X}}$'s to be uncorrelated, and introduce a factor
$f_{\rm sky}$, which denotes the observed fraction of sky. As was
shown in \cite{ours,cutpdf}, for the estimators with the multipole
number $\ell$, the number of degree of freedom reduces to $n_{\rm
eff}=(2\ell+1)f_{\rm sky}$ (instead of $n=2\ell+1$). Thus,
compared to the full sky, the inclusion of cut sky reduces the
degree of freedom in the definition of the estimators
$D_{\ell}^{XX'}$. In this work, we shall discuss the CMB field
with the cut sky factor
\begin{eqnarray}\label{cu}
 f_{\rm sky}=0.65,
\end{eqnarray}
which is suggested by Planck bluebook \cite{Planck}. In all the
following discussion, we should remember to replace $n$ with the
effective degree of freedom  $n_{\rm eff}$, when using the
result in (\ref{mean--variances}) and the pdfs in (\ref{chi2}),
(\ref{wishart}), (\ref{pdf_CT}) or (\ref{joint}).


\section{Four cases to detect RGWs in the CMB \label{s2}}

In this paper, we shall investigate the detection abilities for
the RGWs in the following four cases: ``{\it B}" case, ``{\it CT}" case,
``{\it CTE}" case and ``{\it CTEB}" case, which will be introduced separately
in this section.


\subsection{{\rm ``{\it B}"} case}

The first case is the well-known ``{\it BB}" method. In this case, one
can detect the signal of RGWs only by the observable
$D_{\ell}^{BB}$, which satisfies the $\chi^2$ distribution in Eq.
(\ref{chi2}).

In order to study the determination of cosmic parameters from the
observed data, we shall consider the likelihood function. The
likelihood is a term, customarily, used to call the probability
density function considered a function of an unknown parameter. Up
to a constant, independent of its arguments, the likelihood is
defined as the pdf of the set of the moments $D_{\ell}^{BB}$ given
$C_{\ell}^{BB}$, i.e.
 \begin{eqnarray}\label{blikelihood1}
 \mathcal{L}_{\rm B}\propto \prod_{\ell}f(D_{\ell}^{BB}).
 \end{eqnarray}
Using the pdf in (\ref{chi2}), and considering the effective
degree of freedom $n_{\rm eff}=(2\ell+1)f_{\rm sky}$ in the cut
sky, the likelihood function in (\ref{blikelihood1}) can be
rewritten as
  \begin{eqnarray}\label{blikelihood2}
  -2\ln \mathcal{L}_{\rm B}=\sum_{\ell} n_{\rm eff}\left\{\left(\frac{D_{\ell}^{BB}+N_{\ell}^{BB}W_{\ell}^{-2}}{(\sigma_{\ell}^B)
^2}\right)-\ln\left(\frac{D_{\ell}^{BB}+N_{\ell}^{BB}W_{\ell}^{-2}}{(\sigma_{\ell}
^B)^2}\right)\right\}+C_1,
  \end{eqnarray}
where $C_1$ is the constant for the normalization.  The noise
power spectrum $N_{\ell}^{BB}$ for Planck mission is given by Eq.
(\ref{noises2}).

Since the $B$-polarization can only be generated by the
gravitational waves, the observable $B$-polarization power
spectrum includes a clean information of the gravitational waves.
This is the advantage of ``{\it BB}" method. However, the amplitude of
the $B$-polarization is expected to be very small, which makes the
detection of $B$-polarization quite difficult. In addition, the
signal of RGWs in $B$-polarization can be contaminated by the
$E$-$B$ mixing due to the partial sky coverage \cite{wuran2}, beam
asymmetry \cite{wuran3} and cosmic lensing effect \cite{wuran4}.
These all can degrade the detection ability of the ``{\it BB}" method.


\subsection{{\rm ``{\it CT}"} case}

Different from the ``{\it BB}" method, in the previous work \cite{ours},
we have detailed discussed the ``{\it TE}" method, detecting the signal
of RGWs by the CMB $TE$ power spectrum. In this method, the
amplitude of $C_{\ell}^{TE}$ is two order larger than
$C_{\ell}^{BB}$. Another advantage of this method is that, the
$E$-$B$ mixing, which can occur for some reasons, nearly cannot
influence $TE$ power spectrum. So it cannot degrade of the
detection ability of this method. However, in the previous work
\cite{ours}, we find that, the uncertainty of the $TE$ estimator
$D_{\ell}^{TE}$ is very large, due to the cosmic uncertainty. So
comparing with ``{\it BB}" method, ``{\it TE}" method has not only the larger
signal, but also the larger uncertainty.

In this paper, we shall develop the ``{\it TE}" method by combining the
CMB $TE$ and $TT$ power spectra. In the real observations, the
amplitude of $C_{\ell}^{TT}$ is much larger than that of the other
three power spectra.  So combining the $TE$ and $TT$ power spectra
are expected to be a more effective way to detect RGWs. We denote
it as ``{\it CT}" case. In this case, the likelihood function is
 \begin{eqnarray}\label{ctlikelihood1}
 \mathcal{L}_{\rm CT}\propto \prod_{\ell}f(D_{\ell}^{TE}, D_{\ell}^{TT}).
 \end{eqnarray}
Using the pdf in (\ref{pdf_CT}), this likelihood function can be
rewritten as
 \begin{eqnarray}\label{ctlikelihood2}
 -2\ln \mathcal{L}_{\rm
CT}=\sum_{\ell}\left\{\frac{1}{1-\rho^2_{\ell}}\left(\frac{{z}^2}{x{(\sigma_\ell^E
)^2} }+\frac{{x}}{{(\sigma_\ell^T)^2}} -\frac{{2\rho_{\ell}}
{z}}{{\sigma_\ell^T}{\sigma_\ell^E}}\right) +\ln
\left((1-\rho_{\ell}^{2})(\sigma_{\ell}^T)^{2n_{\rm
eff}}(\sigma_{\ell}^E)^2\right)\right\}+C_2.
 \end{eqnarray}


\subsection{{\rm ``{\it CTE}"} case}
In this case, in addition to the $TE$ and $TT$ power spectra, we
shall include the $E$-polarization power spectrum $C_{\ell}^{EE}$.
By comparing with ``{\it CT}" case, we can investigate the contribution
of  $E$-polarization for the detection of RGWs.  In this case, the
likelihood function is
\begin{eqnarray}\label{ctelikelihood1}
 \mathcal{L}_{\rm CTE}\propto \prod_{\ell}f(D_{\ell}^{TE}, D_{\ell}^{TT},
D_{\ell}^{EE}).
 \end{eqnarray}
Using the pdf in (\ref{wishart}), this likelihood can be written
as
  \begin{eqnarray}\label{ctelikelihood2}
  -2\ln \mathcal{L}_{\rm CTE}=\sum_{\ell}
\left\{\frac{1}{(1-{\rho_\ell^2})}\left(\frac{{x}}{(\sigma_\ell^T)^2}+\frac{{y}}{(
\sigma_{\ell}^E)^2}-\frac{2{\rho_l}
{z}}{{\sigma_{\ell}^T}{\sigma_{\ell}^E}}\right)
  +n_{\rm
eff}\ln\left(4(1-\rho_\ell^2)(\sigma_\ell^T\sigma_\ell^E)^2\right)\right\}+C_3.
  \end{eqnarray}


\subsection{{\rm ``{\it CTEB}"} case}
This case will use all the CMB power spectra, $C_{\ell}^{TE},
C_{\ell}^{TT}, C_{\ell}^{EE}$ and $C_{\ell}^{BB}$, so it is a
combination of ``{\it CTE}" and ``{\it B}". By investigating this case, we can
determine the best constraint of RGWs by the CMB observation. In
this case, the likelihood is
\begin{eqnarray}\label{cteblikelihood1}
 \mathcal{L}_{\rm CTEB}\propto \prod_{\ell}f(D_{\ell}^{TE}, D_{\ell}^{TT},
D_{\ell}^{EE}, D_{\ell}^{BB}),
 \end{eqnarray}
which is the product of $\mathcal{L}_{\rm CTE}$ and
$\mathcal{L}_{\rm B}$, i.e.
\begin{eqnarray}\label{cteblikelihood2}
 \mathcal{L}_{\rm CTEB}=\mathcal{L}_{\rm CTE}\mathcal{L}_{\rm B}.
 \end{eqnarray}


\section{The simulation method and the results\label{s3}}

As the previous work \cite{knox}, in this section, we shall use
the maximum likelihood analysis, based on the simulated data, to
discuss the sensitivities for the detection of RGWs in these four
cases.

Before proceeding on the simulation method, we shall firstly
introduce the background cosmological model. Throughout this
paper, we shall adopt a set of typical cosmological parameters as
follows \cite{typical}:
\begin{eqnarray}\label{background}
h=0.732,~\Omega_b
h^2=0.02229,~\Omega_{m}h^2=0.1277,~\Omega_{k}=0,~\tau_{reion}=0.089.
\end{eqnarray}
Since in this paper, we focus on the detection abilities for the
RGWs, in Sections \ref{s3} and \ref{s4}, we shall only consider
the constraint on the parameter $r$. In Section \ref{s5}, we shall
extend to the constraint on the other three parameters $n_t$,
$A_s$, $n_s$, and discuss their influence on the constraint of
$r$. The extent of the constraints on the cosmological parameters
($h$, $\Omega_b$, $\Omega_m$, $\Omega_k$, $\tau_{reion}$) remains
an open question in this paper. Actually, by the forthcoming
observation of Planck satellite, the constraints on these
cosmological parameters are expected to be very tight. For
example, the constraint on $\Omega_bh^2$ would be $\Delta
\Omega_bh^2=0.00017$, the constraint on $\tau_{reion}$ would be
$\Delta \tau_{reion}=0.005$ \cite{Planck}, which are expected to
have little influence on the determination of RGWs. In all this
paper, we take specific values for cosmological parameters as in
(\ref{background}) and assume that they are perfectly known.


\subsection{The method\label{s3.1}}

In this section, we shall use the maximum likelihood analysis to
investigate the constraint on the cosmological parameters. This
method has been used in the previous work \cite{knox} for the CMB
analysis and in the work \cite{darkenergy} for dark energy
analysis. If we consider the ``{\it B}" case, the steps of the method
can be listed as the follows (the similar steps can also be used
in the ``{\it CT}", ``{\it CTE}" and ``{\it CTEB}" cases):

\textbf{Step 1} We build the pdf of the estimator:
$f(D_{\ell}^{BB})$, which have been given in Eq. (\ref{chi2}).

\textbf{Step 2} According to this pdf, we generate $N$ sets of
random samples
$\{D_{\ell}^{BB}|\ell=2,3,\cdot\cdot\cdot,\ell_{\max}\}$ (we call
each sample as a ``realization"), where the input model has the
parameters ($\hat{r}$, $\hat{n}_t$, $\hat{A}_{s}$, $\hat{n}_s$)
\footnote{ Throughout this paper, the parameters of the input
cosmological model are marked with a $``hat"$ superscript.}.

\textbf{Step 3} We separate the these parameters into two sets:
the first set includes the so-called \emph{unfixed} parameters,
and the second set includes the \emph{fixed} parameters. In this
section, we are only interested in the constraint on the amplitude
of RGWs, so we consider the simplest case, where the only
\emph{unfixed} parameter is the tensor-to-scalar ratio $r$. The
other three parameters, $n_t$, $A_s$ and $n_s$, are all the
\emph{fixed} parameters.  In Section \ref{s5}, we shall discuss
the influence of other parameters on constraint of $r$, so we
shall choose more than one parameters as the \emph{unfixed}
parameters.

\textbf{Step 4} We fix the \emph{fixed} parameters as
their input values and set the \emph{unfixed}
ones as the free parameters. Using Eq. (\ref{blikelihood2}), an
automated search, which uses the numerical technique of simulated
annealing \cite{numerical}, finds the maximum of likelihood
$\mathcal{L}_{\rm B}$ for each realization.

\textbf{Step 5} To measure the certainty with which the
\emph{unfixed} parameters can be determined, we examine the
distribution of the maxima from the simulations.

Evaluation of the likelihood  function on a fine grids of the
\emph{unfixed} parameters shows that the maximum found by the
automated procedure differs negligibly from the true maximum.
Performing $N$ realization allows us to determine the standard
deviations of the \emph{unfixed} parameters with a fractional
error of $(2N)^{-0.5}$. When $N=300$, the fractional error is
$4\%$, and when $N=1000$,  the fractional error is $2\%$.


\subsection{Results}

We apply the simulation method to the ``{\it B}", ``{\it CT}", ``{\it CTE}" and
``{\it CTEB}" cases. The values of $\ell_{\max}$, $N$ and the input
values of parameters are adopted as the follows:
\begin{eqnarray}\label{coefficients}
\ell_{\max}=100,~~N=1000,~~\hat{r}=0.3, ~~\hat{n}_t=0.0,
~~\hat{A}_{s}=2.3\times10^{-9}, ~~\hat{n}_s=1.0.
\end{eqnarray}
Since when considering the Planck instrumental noises, the
contributions of RGWs in the CMB power spectra $C_{\ell}^{XX'}$
are important only at the large scale ($\ell\leq100$) \cite{ours},
we have adopt $\ell_{\max}=100$, i.e. only using the simulated
data in the range $\ell\leq 100$ in the likelihood analysis. For
each case, we generate $(N=)1000$ realization, which makes that we
can determine the standard deviation of the \emph{unfixed}
parameter with a fractional error of $2\%$.

In this section, we only set $r$ as the \emph{unfixed} parameter,
i.e. the spectral indices $n_t$, $n_s$ and the amplitude $A_s$
will be fixed as their input values in the calculation. As
mentioned above, for any two pivot wavenumbers $k_0$ and $k_1$,
the constraints on $r(k_0)$ and $r(k_1)$ are exactly same. So in
the discussion in this section, we shall not differentiate $r$ at
the different pivot wavenumbers.

FIG.1 presents the distribution of the maxima $r_p$ in each case.
We find in all of these realization, the values of $r_p$ are close
to the input value $\hat{r}=0.3$. In ``{\it B}" case, we find the
average value of $r_p$ is $\overline{r_p}=0.305$. In ``{\it CT}" case,
we have $\overline{r_p}=0.298$. In ``{\it CTE}" case, we have
$\overline{r_p}=0.303$, and in ``{\it CTEB}" case, we have
$\overline{r_p}=0.304$. These four average values are all equal to
the input value within $1.7\%$ ($<2\%$) and hence  there is no
evidence for bias.

In the different cases, the diffusion of $r_p$ is different. The
standard deviation of $r_p$ in ``{\it B}" case is $\Delta r_p=0.067$, so
we can conclude that an experiment of this type can determine $r$
with $\Delta r=0.067\pm0.001$. In ``{\it CT}" case, the standard
deviation  is $\Delta r_p=0.078$, which is $16.4\%$ larger than
that in ``{\it B}" case.  So we conclude that ``{\it CT}" is a little less
sensitive than ``{\it B}" for the constraint of $r$. In ``{\it CTE}" case, the
standard deviation is $\Delta r_p=0.070$, which is $10.3\%$
smaller than that in ``{\it CT}" case. So including the $E$-polarization
data, the constraint on $r$ becomes tighter. However, it is also
$3.0\%$ larger than the value of $\Delta r_p$ in ``{\it B}" case.  In
``{\it CTEB}" case, we have $\Delta r_p=0.047$. which is much smaller
than the others. So we conclude that, taking into account all the
simulated data, the constraint on the parameter $r$ can be much
improved. Comparing with ``{\it B}", the uncertainty of $r$ is reduced
by $29.8\%$, and comparing with ``{\it CT}", the uncertainty is reduced
by $39.7\%$.

\begin{figure}[t]
\centerline{\includegraphics[width=14cm,height=8cm]{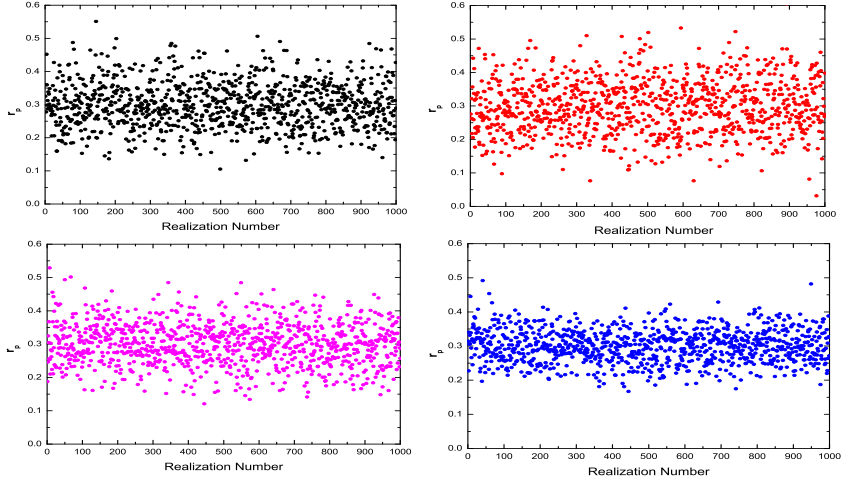}}
\caption{ The distribution of $r_p$ in the 1000 realization. The
black dots (upper left panel) denote the result in ``{\it B}" case, the
red dots (upper right panel) are the result in ``{\it CT}" case, the
magenta dots (lower left panel) are the result in ``{\it CTE}" case, and
the blue dots (lower right panel) denote the result in ``{\it CTEB}"
case. In all these panels, we have considered one free parameter
$r$ in the likelihood analysis. The input simulated data are up to
$\ell_{\max}=100$, and the input value is
$\hat{r}=0.3$.}\label{figure1}
\end{figure}

\begin{figure}[t]
\centerline{\includegraphics[width=12cm,height=8cm]{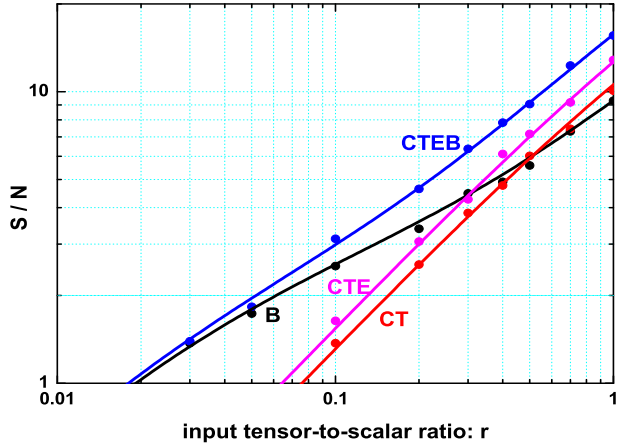}}
\caption{The value of $S/N$ depends on the input value $\hat{r}$.
The black, red, magenta and blue dots (lines) are the simulation
(analytic approximation) results in ``{\it B}" case, ``{\it CT}" case, ``{\it CTE}"
case and ``{\it CTEB}" case, respectively. }\label{figure2}
\end{figure}

Similar to the discussion in our previous work \cite{ours}, in
order to describe the detection abilities for the RGWs, we define
the signal-to-noise ratio
\begin{eqnarray}\label{snr}
S/N\equiv \hat{r}/\Delta r_p, \end{eqnarray} where $\hat{r}$ is
the input value of $r$. We can determine this quantity with
different input $\hat{r}$. For each input, we generate $1000$
realization, and calculate the quantities $\overline{r_p}$,
$\Delta r_p$ and $S/N$. The value of $S/N$ as a function of
$\hat{r}$ are shown in FIG.2.  From this figure, let us firstly
investigate the detection abilities in the four cases. We find, in
``{\it B}" case, the parameter $r$ can be determined at 2$\sigma$ level
when $\hat{r}>0.06$.  In ``{\it CT}" case,  $r$ can be determined at
2$\sigma$ level when $\hat{r}>0.16$.  In ``{\it CTE}" case,  $r$ can be
determined at 2$\sigma$ level when $\hat{r}>0.13$, and in``{\it CTEB}"
case,  $r$ can be determined at 2$\sigma$ level when
$\hat{r}>0.05$.

From FIG.2, We can also compare the sensitivities in the different
cases by the values of $S/N$.  Comparing ``{\it CT}" and ``{\it B}", we find
the former one is more sensitive when $\hat{r}$ is large, and the
latter one is more sensitive when $\hat{r}$ is small. ``{\it CT}" is
more sensitive than ``{\it B}" when $\hat{r}>0.5$. ``{\it CTE}" is more
sensitive than ``{\it B}", when $\hat{r}>0.3$.  In order to investigate
the contribution of $E$-polarization on the detection of RGWs, we
can compare the values of $S/N$ in ``{\it CT}" and ``{\it CTE}" cases. From
the FIG.2, we find the quantity $S/N$ in ``{\it CTE}" case is always
$15\%$ larger than that in ``{\it CT}" case. So considering the
$E$-polarization, the constraint on $r$ can be improved for any
$\hat{r}$. From the FIG.2, we also find that, as the combination
of ``{\it CTE}" and ``{\it B}", ``{\it CTEB}" is more sensitive than the other three
cases. When $\hat{r}$ is small, the sensitivity in ``{\it CTEB}" case is
close to that in ``{\it B}" case, since in this case, the sensitivity of
``{\it CTE}" is very weak. When $\hat{r}$ is large, the sensitivity in
``{\it CTEB}" is close to that in ``{\it CTE}" case.

\begin{table}
\caption{The mean values and standard deviations of $r_p$. In the
likelihood analysis, we have considered one free parameter $r$. }
\begin{center}
\label{1para}
\begin{tabular}{|c|c|c|c|c|c|}
  \hline
    input $\ell_{\max}$ & output parameter& B & CT& CTE & CTEB \\
  \hline
  $100$  & $\overline{r_p}\pm\Delta r_p$ &   $0.305\pm0.067$ & $0.298\pm0.078$ &
$0.303\pm0.070$  & $0.304\pm0.047$ \\
  \hline
   $500$ & $\overline{r_p}\pm \Delta r_p$ &  $0.302\pm0.067$ & $0.301\pm0.080$ &
$0.302\pm0.066$ & $0.302\pm0.047$\\
  \hline
\end{tabular}
\end{center}
\end{table}

We have also applied the simulation method to another condition:
the input quantities are all exactly same with those in Eq.
(\ref{coefficients}), except the value of $\ell_{\max}$. In this
case we adopt $\ell_{\max}=500$, i.e. the simulated data
$D_{\ell}^{XX'}$ in the range $\ell\leq500$ are used for the
likelihood analysis. In Table \ref{1para} we summarize the output
values $\overline{r_p}\pm\Delta r_p$ in ``{\it B}", ``{\it CT}", ``{\it CTE}",
``{\it CTEB}" cases, where $\hat{r}=0.3$ is used. We find that, the
results in this condition is very close to those in the condition
with $\ell_{\max}=100$. This result testifies that, when
considering the Planck instrumental noises, the contribution of
RGWs in the CMB power spectra $C_{\ell}^{XX'}$ are important only
at the large scale ($\ell\leq100$).

\begin{figure}[t]
\centerline{\includegraphics[width=12cm,height=8cm]{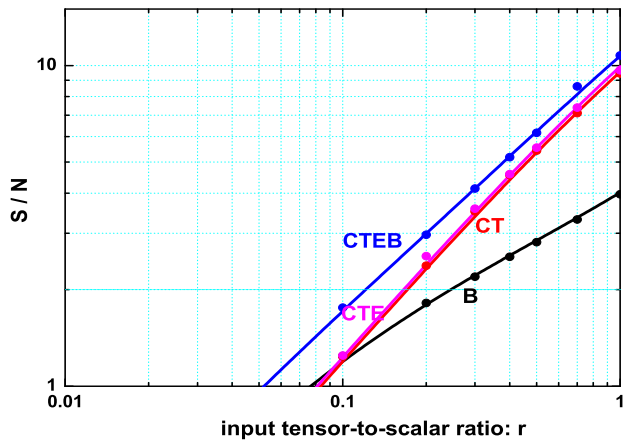}}
\caption{The same graph as in FIG.\ref{figure2}, the only
difference is that, in this figure we have assumed the realistic
noise power spectra $N_{\ell}^{XX}$ are 4 times larger than the
Planck instrumental noises.}\label{figure3}
\end{figure}

To this point we have assumed that the noise power spectra only
come from the Planck instrumental noise. However, synchrotron and
bremsstrahlung radiation, thermal emission from cold dust, and
unsolved extragalactic sources also contribute to the anisotropy
and polarization of radiation. These contaminations can enlarge
the effective ``noises" of CMB power spectra
\cite{knox,5map,noises22}.  In order to estimate the effect of
these contaminations on the constraint of $r$, in this paper, we
only simply assume the foreground will degrade the noise $a_{\ell
m}^{X}(n)$ by a factor $2$. Therefore, we take into account the
effect of foreground contaminations by simply increasing
$N_{\ell}^{XX}$ to $4N_{\ell}^{XX}$.  In this case, by the exactly
same steps as the previous discussion, we recalculate the
signal-to-noise ratio $S/N$ by the simulation method, where
different input values $\hat{r}$ are considered. The quantity
$S/N$ as a function of $\hat{r}$ in four cases are shown in FIG.3.
Let us firstly discuss the results in ``{\it B}" case.  Comparing with
the results in FIG.2, we find the detection ability in ``{\it B}" case
is much decreased. Only when $\hat{r}>0.25$, the signal of RGWs
can be detected in 2$\sigma$ level. However, in ``{\it CT}" and ``{\it CTE}"
cases, the results are similar with those in the previous
condition with only instrumental noises (FIG.2). Comparing the
sensitivities in ``{\it CT}" and ``{\it CTE}" cases, we find the difference is
very small, which suggests that the contribution of
$E$-polarization for the detection of RGWs is negligible. In
``{\it CTEB}" cases, when $\hat{r}>0.12$, the signal of RGWs can be
detected in 2$\sigma$ level.

In the previous work \cite{ours}, we found that WMAP5 $TE$ data
induces the best-fit model with $r\simeq0.48$. From FIGs.
\ref{figure2} and \ref{figure3}, we find that RGWs  with $r=0.48$
will be presented nearly at $9\sigma$ level in ``{\it CTEB}" case, when
the Planck instrumental noises are considered. If the assumed
realistic noises are considered, it will be presented at $6\sigma$
level. These are all much better than the results in ``{\it TE}" and
``{\it BB}" methods \cite{ours}.


\section{Analytic approximation of $S/N$\label{s4}}

In Section \ref{s3}, using the signal-to-noise ratio $S/N$
calculated by the simulated data, we have investigated the
detection abilities for the RGWs in four cases (``{\it B}", ``{\it CT}",
``{\it CTE}", ``{\it CTEB}" cases). In order to better understand this
signal-to-noise ratio and get an intuitive feel for the results in
Section  \ref{s3}, in this section we shall give a simple analytic
approximation of the signal-to-noise ratio. Similar to the
discussion in Section \ref{s3}, in the analytic approximation, we
will also be interested in a single \emph{unfixed} parameter, the
tensor-to-scalar ratio $r$. Other parameters ($n_t$, $A_s$ and
$n_s$) and background parameters ($h$, $\Omega_b$,
$\cdot\cdot\cdot$) are all assumed to be exactly known.


\subsection{Analytic approximation of the likelihood functions}

In order to present the analytic expression of the signal-to-noise
ratio, we need to express the likelihoods as the simple functions
of variable $r$. We notice that the exact pdfs
in Eqs. (\ref{chi2}), (\ref{wishart}), (\ref{pdf_CT}),
(\ref{joint}), are all very close to the Gaussian function,
especially when $\ell\gg 1$ (see for instant \cite{ours,
wishart2}). Based on the Gaussian approximation of these pdfs, the
likelihood functions in Eqs. (\ref{blikelihood1}),
(\ref{ctelikelihood1}), (\ref{ctlikelihood1}),
(\ref{cteblikelihood1}) can be simplified as
(\ref{blikelihood-ana2}), (\ref{ctelikelihood-ana2}),
(\ref{ctlikelihood-ana2}), (\ref{cteblikelihood-ana2}) (see
Appendix \ref{appendix} for the details), which can be rewritten
in a unified form as follows:
 \begin{eqnarray} \label{unified}
-2\ln\mathcal{L}(r)=\sum_{\ell=2}^{\ell_{\max}}\sum_{XX'}\left[\frac{C_{\ell}^{XX'
}-D_{\ell}^{XX'}(\hat{r})}{\Delta
D_{\ell}^{XX'}(\hat{r})}\right]^2.\end{eqnarray} In ``{\it B}" case, we
have $XX'=BB$; in ``{\it CT}" case, we have $XX'=TE,TT$; in ``{\it CTE}" case,
$XX'=TE,TT,EE$; and in ``{\it CTEB}" case, we have $XX'=TE,TT,EE,BB$. In
each case, the likelihood function $\L$ depends on the variable  $r$
only by the power spectra $C_{\ell}^{XX'}$.

In general, ignoring the possible contribution from the (vector)
rotational perturbations, the CMB power spectra $C_{\ell}^{XX'}$
can be presented as a sum of two contributions: density
perturbations and gravitational waves:
\begin{eqnarray} \label{sum-C}
C_{\ell}^{XX'}(r)=C_{\ell}^{XX'}(dp)+C_{\ell}^{XX'}(gw),\end{eqnarray}
where $C_{\ell}^{XX'}(dp)$ and $C_{\ell}^{XX'}(gw)$ are the
contributions of density perturbations and gravitational waves,
respectively. We should remember $C_{\ell}^{BB}(dp)=0$. In
the likelihood analysis, we have fixed the parameters
$n_t$, $A_s$, $n_s$ as their input values, and only considered a
single free parameter $r$. $C_{\ell}^{XX'}(gw)$ depends
on the variable  $r$, which can be written as
\begin{eqnarray}
\label{sum-C2}C_{\ell}^{XX'}(gw)=\left(\frac{r}{\hat{r}}\right)\hat{C}_{\ell}^{XX'
}(gw), \end{eqnarray} where $\hat{C}_{\ell}^{XX'}(gw)$ are the
power spectra $C_{\ell}^{XX'}(gw)$ at $r=\hat{r}$.  Inserting
(\ref{sum-C2}) in (\ref{sum-C}), we get
\begin{eqnarray} \label{sum-C3}
C_{\ell}^{XX'}(r)=C_{\ell}^{XX'}(dp)+\left(\frac{r}{\hat{r}}\right)\hat{C}_{\ell}^
{XX'}(gw).\end{eqnarray}

Now, let us return to the likelihood function. Inserting
(\ref{sum-C3}) into Eq. (\ref{unified}), we obtain that
\begin{eqnarray} \label{unified2}
-2\ln\mathcal{L}(r)=\sum_{\ell=2}^{\ell_{\max}}\sum_{XX'}\alpha^{XX'}_{\ell}\left(
\frac{r}{\hat{r}}-\beta^{XX'}_{\ell}\right)^2,\end{eqnarray} where
the quantities $\alpha_{\ell}^{XX'}$ and $\beta_{\ell}^{XX'}$ are
defined by
\begin{eqnarray} \label{alpha-beta} \alpha^{XX'}_{\ell}\equiv
\left(\frac{\hat{C}_{\ell}^{XX'}(gw)}{\Delta
D_{\ell}^{XX'}(\hat{r})}\right)^2,~~ \beta^{XX'}_{\ell}\equiv
\frac{D_{\ell}^{XX'}(\hat{r})-C_{\ell}^{XX'}(dp)}{\hat{C}_{\ell}^{XX'}(gw)}.
\end{eqnarray}
After straight forward manipulations, the expression
(\ref{unified2}) can be rewritten as the following form
\begin{eqnarray}\label{unified3}
-2\mathcal{L}(r)=(\sum_{\ell=2}^{\ell_{\max}}\sum_{XX'}\alpha_{\ell}^{XX'})
\left(\frac{r}{\hat{r}}-\frac{\sum_{\ell=2}^{\ell_{\max}}\sum_{XX'}
\alpha_{\ell}^{XX'}\beta_{\ell}^{XX'}}{\sum_{\ell=2}^{\ell_{\max}}\sum_{XX'}\alpha_{\ell}^{XX'}}\right)^
2+C',
\end{eqnarray}
where the separate part $C'$ is independent of the variable  $r$. In
the following, based on this formula, we shall discuss the
signal-to-noise ratio $S/N$.


 \subsection{Analytic approximation of $S/N$}

Now, let us investigate the likelihood in (\ref{unified3}). First,
we shall discuss the peak of the likelihood function. We notice
that this likelihood is a ``Gaussian
form function" of the variable  $r$.  The maximum is at $r_p$, which is
\begin{eqnarray}\label{rp}
r_p=\frac{\sum_{\ell=2}^{\ell_{\max}}\sum_{XX'}\alpha_{\ell}^{XX'}\beta_{\ell}^{XX
'}}{\sum_{\ell=2}^{\ell_{\max}}\sum_{XX'}\alpha_{\ell}^{XX'}}\hat{r}.\end{eqnarray}
From the definition of $\alpha_{\ell}^{XX'}$ and
$\beta_{\ell}^{XX'}$ in Eq. (\ref{alpha-beta}), we find the value
of $r_p$ not only depends on the values of
$C_{\ell}^{XX'}(\hat{r})$, the input model, and $N_{\ell}^{XX}$,
the noise power spectra, but also depends on the values of
$D_{\ell}^{XX'}$, the simulated data. So for any two realization,
even if generated by the exactly same input model and noises, they
have the different values of $r_p$.

From the likelihood in Eq. (\ref{unified3}), we can also obtain the
spread of the likelihood $\Delta r$, which is
\begin{eqnarray}\label{delta-r}
\Delta
r=\frac{\hat{r}}{\left(\sum_{\ell=2}^{\ell_{\max}}\sum_{XX'}\alpha_{\ell}^{XX'}\right)^{1/2}}.\end{eqnarray}
The spread of the likelihood only depends on the values of
$C_{\ell}^{XX'}(\hat{r})$, the input model, and $N_{\ell}^{XX}$,
the noise power spectra. So we get the conclusion, for any two
realization, as long as they have the same input model and noises,
they have the same $\Delta r$, the spread of the likelihood
function.

In the previous work \cite{ours}, we have defined the
signal-to-noise ratio $\widetilde{S/N}$ as
\begin{eqnarray} \label{old-snr} \widetilde{S/N}\equiv {\hat{r}}/{\Delta
r}.\end{eqnarray} Note that, in order to distinguish from $S/N$
defined in (\ref{snr}), we denote the signal-to-noise ratio in our
previous work as $\widetilde{S/N}$. In the following, we will find
these two definitions have the exactly same values. Using the
formula in (\ref{delta-r}) and the definition of
$\alpha_{\ell}^{XX'}$ in (\ref{alpha-beta}), we obtain
\begin{eqnarray} \label{old-snr2}
\widetilde{S/N}=\sqrt{\sum_{\ell=2}^{\ell_{\max}}\sum_{XX'}\left(\frac{\hat{C}_{\ell}^{XX'}(gw)}{\Delta
D_{\ell}^{XX'}(\hat{r})}\right)^2}.\end{eqnarray} This is the
finial analytic result of the quantity $\widetilde{S/N}$, which
depends on the input power spectra $C_{\ell}^{XX'}(\hat{r})$ and
the noise power spectra $N_{\ell}^{XX'}$. In the work
\cite{jaffe}, by the Fisher Matrix analysis, the authors have
obtained a same result as Eq. (\ref{old-snr2}) in the case of
$XX'=BB$ (the result in ``{\it B}" case). The results in Eqs.
(\ref{rp}), (\ref{delta-r}) and (\ref{old-snr2}) describe the
constraint on the tensor-to-scalar ratio $r$, based on one set of
simulated data
$\{D_{\ell}^{XX'}|\ell=2,3,\cdot\cdot\cdot,\ell_{\max}\}$.

However, in our discussion in Section \ref{s3}, we have considered
another case. In the simulation method, based on a same input
cosmological model and noises, we have randomly generated $N$
($N\gg1$) realization.  For each realization, we can obtain a
maximum of likelihood $r_p$. From these $r_p$, we have calculated
the mean value and standard deviation of $r_p$. From the
simulation, we find the mean value $\overline{r_p}$ is close to
the input value $\hat{r}$, and the standard deviation $\Delta r_p$
stands for the uncertainty of the parameter $r$, in the likelihood
analysis. Now, we shall prove these in the analytic approximation.

In the analytic approximation, from Eq. (\ref{rp}) we can also
calculate the values of $\overline{r_p}$ and $\Delta r_p$. The
mean value $\overline{r_p}$ is
\begin{eqnarray} \label{rp-mean1}\overline{r_p} =
\frac{\sum_{\ell=2}^{\ell_{\max}}\sum_{XX'}\alpha_{\ell}^{XX'}\langle
\beta_{\ell}^{XX'}\rangle}{\sum_{\ell=2}^{\ell_{\max}}\sum_{XX'}\alpha_{\ell}^{XX'
}}\hat{r}.\end{eqnarray} Considering the definition of
$\beta_{\ell}^{XX'}$ in Eq. (\ref{alpha-beta}), the relation
$\langle D_{\ell}^{XX'}(\hat{r})\rangle=C_{\ell}^{XX'}(\hat{r})$
and  the Eq. (\ref{sum-C3}), we can obtain that
$\langle\beta_{\ell}^{XX'}\rangle=1$. Inserting it into Eq.
(\ref{rp-mean1}), we get a relation \begin{eqnarray}
\label{rp-mean2} \overline{r_p} = \hat{r}. \end{eqnarray} This
relation suggests that, the mean value of $r_p$ is equal to the
input value $\hat{r}$, which is consistent with the simulation
result in Section \ref{s3}.

We can also discuss the standard deviation of $r_p$, which is
defined by $\Delta
r_p\equiv\sqrt{\overline{r_p^2}-{\overline{r_p}}^2}$. Using
Eq. (\ref{rp}) and the relation $\langle
D_{\ell}^{XX'}(\hat{r})\rangle=C_{\ell}^{XX'}(\hat{r})$, after a
straight forward manipulations, we obtain that
\begin{eqnarray} \label{delta-rp}\Delta r_p
=\frac{\hat{r}}{\left(\sum_{\ell=2}^{\ell_{\max}}\sum_{XX'}\alpha_{\ell}^{XX'}\right)^{1/2}}.\end{eqnarray}
Comparing (\ref{delta-rp}) with (\ref{delta-r}), we find $\Delta
r_p=\Delta r$. By the formula (\ref{delta-rp}), we can discuss the
signal-to-noise ratio $S/N$, defined by Eq. (\ref{snr}). Taking
into account the definition of $\alpha_{\ell}^{XX'}$ in
Eq. (\ref{alpha-beta}), we can write the signal-to-noise ratio as
\begin{eqnarray}\label{snr2}
S/N=\widetilde{S/N}=\sqrt{\sum_{\ell=2}^{\ell_{\max}}\sum_{XX'}\left(\frac{\hat{C}
_{\ell}^{XX'}(gw)}{\Delta D_{\ell}^{XX'}(\hat{r})}\right)^2},
\end{eqnarray}
which depends on the input power spectra $C_{\ell}^{XX'}(\hat{r})$
and the noise power spectra $N_{\ell}^{XX'}$. In Eq. (\ref{snr2}),
we should remember that, $XX'=BB$ in ``{\it B}" case, $XX'=TE,TT$ in
``{\it CT}" case, $XX'=TE,TT,EE$ in ``{\it CTE}" case and $XX'=TE,TT,EE,BB$ in
``{\it CTEB}" case. From the expression in (\ref{snr2}), we find that,
the two definitions of signal-to-noise ratio, $S/N$ and
$\widetilde{S/N}$ have the same values. They all stand for the
detection abilities for the RGWs.

Using Eq. (\ref{snr2}), and taking into account the corresponding
noises power spectra $N_{\ell}^{XX'}$, in FIG.\ref{figure2} and
\ref{figure3}, we have plotted the quantities $S/N$ as a function
of $\hat{r}$ in ``{\it B}", ``{\it CT}", ``{\it CTE}", ``{\it CTEB}" cases. We find they
are all exactly consistent with the simulation results.


\subsection{Understanding the analytic approximation of $S/N$}

Now, let us investigate the approximation formula (\ref{snr2}),
which can be rewritten as \begin{eqnarray} \label{snr3}
(S/N)^2=(\widetilde{S/N})^2={\sum_{\ell=2}^{\ell_{\max}}\sum_{XX'}\left(\frac{\hat
{C}_{\ell}^{XX'}(gw)}{\Delta
D_{\ell}^{XX'}(\hat{r})}\right)^2}.\end{eqnarray} In this
expression, $\hat{C}_{\ell}^{XX'}(gw)$ is the contribution of RGWs
to the total power spectra, which determines the strength of the
signal of RGWs. $\Delta D_{\ell}^{XX'}$ is the uncertainty of the
estimator, which serves as the corresponding `noises'.  So the
total signal-to-noise ratio $S/N$ (or $\widetilde {S/N}$) is
determined by the sum of the ratios between RGWs signal and the
corresponding `noise' at every multipole and $XX'$. This result is
consistent with that in ``{\it TE}" method, which has been obtained in
our previous work \cite{ours}.

From Eq. (\ref{snr3}), we can discuss the contribution of each
$\ell$ to the total signal-to-noise ratio. We define the
signal-to-noise ratio at the individual multipole $\ell$,
$(S/N)_{\ell}$ as below:
\begin{eqnarray} \label{snrl} (S/N)_{\ell}^2\equiv
\sum_{XX'}\left(\frac{\hat{C}_{\ell}^{XX'}(gw)}{\Delta
D_{\ell}^{XX'}(\hat{r})}\right)^2.\end{eqnarray} Thus the total signal-to-noise ratio can be
written as the following sum:
\begin{eqnarray} \label{snr4}
(S/N)^2=(\widetilde{S/N})^2=\sum_{\ell=2}^{\ell_{\max}}(S/N)^2_{\ell}.\end{eqnarray}

\begin{figure}[t]
\centerline{\includegraphics[width=15cm,height=7cm]{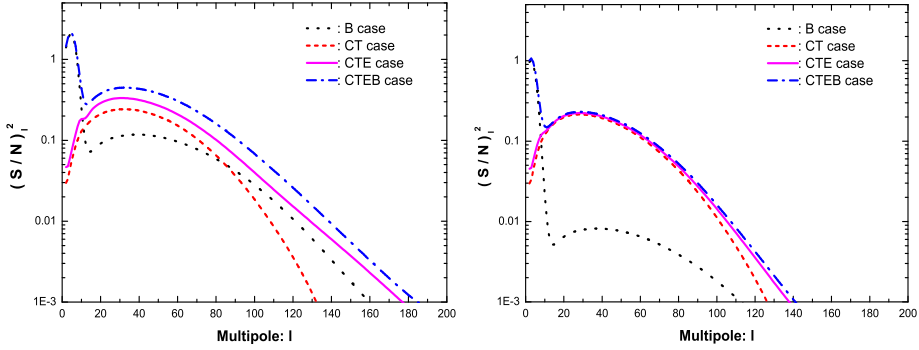}}
\caption{ This figure shows the individual signal-to-noise ratio
$(S/N)_{\ell}^2$ as a function of $\ell$. The presented in left
panel is the result with the Planck instrumental noises, and the
presented in right panel is the result in the case where we assume
the realistic noise power spectra $N_{\ell}^{XX}$ are 4 times
larger than the Planck instrumental noises. }\label{figure4}
\end{figure}

Let us discuss the quantity $(S/N)_{\ell}^2$.  Taking into account
the noise power spectra $N_{\ell}^{XX'}$, and adopt the input
value $\hat{r}=0.3$, in FIG. \ref{figure4} we plot the quantity
$(S/N)_{\ell}^2$ as a function of $\ell$. Let us firstly focus on
the lines in ``{\it B}" case (black lines). We find that
$(S/N)_{\ell}^2$ is sharply peaked at $\ell<10$. As mentioned in
our previous work \cite{ours}, the main contribution in ``{\it B}" case
comes from the signal in the range $\ell<10$. Especially, by
comparing the line in left panel (Planck instrumental noises are
considered) with the right panel (noise power spectra are 4 times
larger than Planck instrumental noises), we find when increasing
the noise power spectra, the value of $(S/N)_{\ell}^2$ reduces by
a factor $2$, at the scale $\ell<10$. However, at the scale
$\ell>10$, the value of $(S/N)_{\ell}^2$ reduces by a factor $10$.
So we get the conclusion, when increasing the noise power spectra,
in ``{\it B}" case, the contribution from $\ell<10$ becomes more and
more dominant.  Since in the range $\ell<10$, the $BB$ power
spectra $C_{\ell}^{BB}$ are mainly generated by the cosmic
reioniztion \cite{a11,a12}, the sensitivity in ``{\it B}" case strongly
depends on the cosmic reionization process.

\begin{figure}[t]
\centerline{\includegraphics[width=15cm,height=7cm]{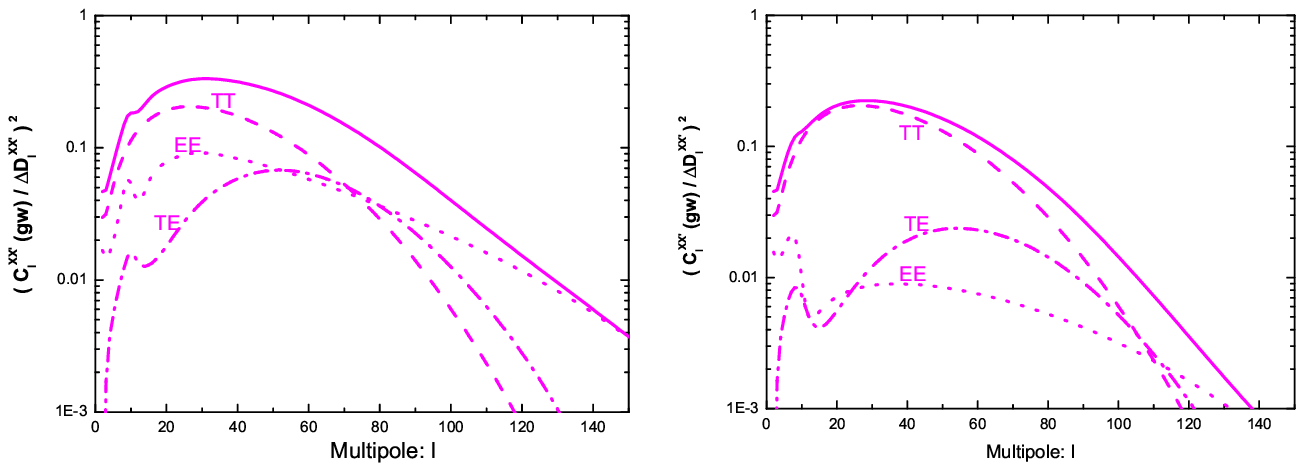}}
\caption{ This figure shows three quantities of
$(\hat{C}_{\ell}^{XX'}(gw)/\Delta D_{\ell}^{XX'})^2$
($XX'=TE,TT,EE$) as a function of $\ell$. As the combination of
these three components,  the individual signal-to-noise ratio
$(S/N)_{\ell}^{2}$ in ``{\it CTE}" case are plotted in solid lines. In
the left panel, we have considered the Planck instrumental noises,
and in the right panel, we have assumed the realistic noise power
spectra are 4 time larger than the Planck instrumental noises.
}\label{figure41}
\end{figure}

Let us turn to the the quantity $(S/N)_{\ell}^2$ in ``{\it CT}" and
``{\it CTE}" cases, which are plotted in red (dashed) and magenta (solid) lines in
FIG.\ref{figure4}.  We find that, in both panels of
FIG.\ref{figure4}, the quantities $(S/N)_{\ell}^2$ in ``{\it CT}" and
``{\it CTE}" cases, are all peaked at $\ell\sim 30$. Among the total
$S/N$, the main contribution comes from the intermedial range
$10<\ell<70$. In the very large scale $\ell<10$ and the small
scale $\ell>70$, the values of $(S/N)_{\ell}^2$ are all very
small. Their contributions to the total $S/N$ are negligible.
Since $(S/N)_{\ell}^2$ is very small in the scale $\ell<10$, the
influence of cosmic reionization on the detection abilities are
also not obvious, which is same with the ``{\it TE}" method, but
different from the ``{\it BB}" method. By comparing the solid and dashed lines in the left panel with the corresponding lines in the right panel, 
we find that, increasing the noises, the values
of $(S/N)_{\ell}^2$ in ``{\it CT}" and ``{\it CTE}" cases have no obvious
change, which induces that the total $S/N$ in ``{\it CT}" and ``{\it CTE}"
cases have no obvious change (see FIG.\ref{figure2} and
\ref{figure3}).

From Eq. (\ref{snrl}), we find, in ``{\it CTE}" (or ``{\it CT}") case, the
quantity $(S/N)_{\ell}^2$ is a simple sum of the portions with
$XX'=TE$, $TT$ and $EE$. So we can also discuss their
contributions to the quantity $(S/N)_{\ell}^2$ separately. In
FIG.\ref{figure41}, we plot these three portions of
$(S/N)_{\ell}^2$ in dash-dotted lines, dashed lines, and dotted
lines.  In this figure, the solid lines denote the sum of these
three portions, which exactly corresponds to the magenta lines in
FIG.\ref{figure4}. We find, when considering the Planck
instrumental noises, these three portions are close to each. In
the range $10<\ell<70$, the largest contribution comes from the
component $XX'=TT$, which is 2 or 3 times larger than the
components $XX'=TE$ and $XX'=EE$.  However, in the case with large
noises (right panel in FIG.\ref{figure41}), the contributions of
the component $XX'=EE$ rapidly decreases, and becomes negligible
among the quantity $(S/N)_{\ell}^2$, which induces that the value
of the quantity $(S/N)_{\ell}^2$ in ``{\it CTE}" case are very close to
that in ``{\it CT}" case (see the right panel of FIG.\ref{figure4}).  So
the total $S/N$ in ``{\it CTE}" and ``{\it CT}" cases are also very close to
each other (see FIG.\ref{figure3}).

We can also discuss the quantity $(S/N)_{\ell}^2$ in ``{\it CTEB}" case,
which are plotted in blue (dash-dotted) lines in FIG.\ref{figure4}. From Eq.
(\ref{snrl}), we find the quantity $(S/N)_{\ell}^2$ in ``{\it CTEB}"
case is a simple sum of $(S/N)_{\ell}^2$ in ``{\it CTE}" case and in
``{\it B}" case. In the range of $\ell<10$, the quantity
$(S/N)_{\ell}^2$ in ``{\it CTEB}" case is close to that in ``{\it B}" case,
and in the range of $\ell>10$, it is close to  that in ``{\it CTE}"
case.


\section{Effects of free parameters: $n_t$ and $A_s$, $n_s$ \label{s5}}

In the previous sections, by both the simulation and the analytic
approximation, we have discussed the constraints of the
tensor-to-scalar ratio $r$ in ``{\it B}", ``{\it CT}", ``{\it CTE}", ``{\it CTEB}" cases.
However, in the real detection, we always have to constrain all
the cosmic parameters, including $h$ the Hubble parameter,
$\Omega_{b}$ the baryon density, $\Omega_m$ the matter density,
$\Omega_{k}$ the spatial curvature, $\tau_{reion}$ the
reionization optical depth. The parameters also include the scalar
spectrum parameters: $A_{s}$ the amplitude of scalar spectrum and
$n_s$ the scalar spectral index, and the tensor spectrum
parameters: $r$ the tensor-to-scalar ratio and $n_t$ the tensor
spectral index.

As mentioned in Section \ref{s3}, in all this paper, we shall not
consider the constraints of the background cosmological
parameters, and assume they have been exactly determined.  In the
likelihood analysis in Sections \ref{s3} and \ref{s4}, we have
considered the case with only one free parameter $r$. The other
parameters $n_t$, $A_s$ and $n_s$ are all fixed as their input
values. Based on this assumption, we have discussed the constraint
on $r$ in ``{\it B}", ``{\it CT}", ``{\it CTE}", ``{\it CTEB}" cases. Thus a question
arises,  if the parameters, $n_t$, $A_s$, $n_s$ are also set free
in the likelihood analysis, whether they can influence the
constraint on the parameter $r$.

In this section, by the simulation method, we shall answer this
question. In the Section \ref{s5.1}, we shall discuss the
constraints on the free parameters, $r$ and $n_t$, and investigate
the effect of $n_t$ on the constraint of $r$. In Section
\ref{s5.2}, we shall consider the free parameters $r$, $n_t$,
$A_{s}$ and $n_s$, and investigate the effects of $A_{s}$ and
$n_s$ on the constraint of $r$.


\subsection{Effect of the free parameter $n_t$\label{s5.1}}

\begin{figure}[t]
\centerline{\includegraphics[width=16cm,height=8cm]{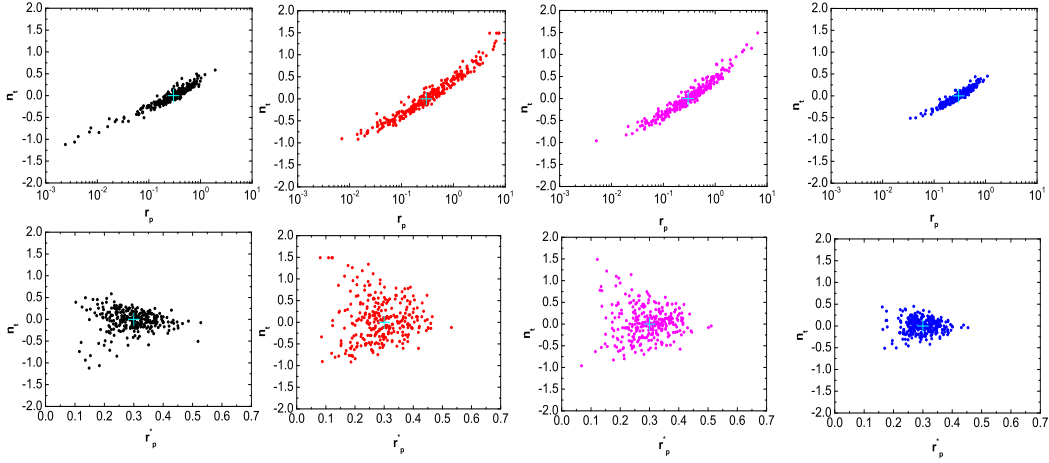}}
\caption{The locations of the maxima from 300 realization
projected into $n_{t}-r_p$ (upper panels), and $n_{t}-r_p^*$
(lower panels) planes. Black (red, magenta, blue) dots denote the
results in ``{\it B}" (``{\it CT}", ``{\it CTE}", ``{\it CTEB}") case. In all these
graphs, we have considered two free parameters ($r$, $n_t$) in the
likelihood analysis. The input simulated data are up to
$\ell_{\max}=100$, and the sign $``+"$ denotes the input values of
the parameters. }\label{figure7}
\end{figure}

Here, we shall use the simulation method described in Section
\ref{s3.1}. We choose the parameters $r$ and $n_t$ as the
\emph{unfixed} parameters, $A_s$ and $n_s$ as the \emph{fixed}
parameters. The values of $\ell_{\max}$, $N$ and the input values
of parameters are adopted as the follows:
\begin{eqnarray}\label{coefficients2}
\ell_{\max}=100,~~N=300,~~\hat{r}=0.3, ~~\hat{n}_t=0.0,
~~\hat{A}_{s}=2.3\times10^{-9}, ~~\hat{n}_s=1.0.
\end{eqnarray}
The background cosmological parameters are adopt as in Eq.
(\ref{background}). We consider the Planck instrumental noises and
Planck window function, which are given in Eqs.
(\ref{noises}-\ref{noises2}). $N=300$ suggests that our following
simulation results ($\overline{r_p}$, $\Delta r_p$,
$\overline{n_t}$, $\Delta n_{t}$) have $4\%$ statistical error.

As mentioned above ($n_s=1$ and $n_t=0$), for any two different pivot wavenumbers $k_0$
and $k_1$,  the tensor-to-scalar ratio $r(k_0)$ and $r(k_1)$ have
the different constraints, due to the free parameter $n_t$.
Although they have the same input values $\hat{r}=0.3$, due to the
formula in (\ref{k1k0}) and $\hat{n}_t-\hat{n}_s+1=0$. As the
first step, in the likelihood analysis, we choose the pivot
wavenumber
\begin{eqnarray}\label{005} k_0=0.05{\rm Mpc}^{-1}. \end{eqnarray} Presented in
FIG.\ref{figure7} (upper panels) shows the maxima projected into
$n_{t}-r_p$ plane from $300$ realization. First, we discuss the
``{\it B}" case. The result is $\overline{r_p}\pm \Delta
r_p=0.351\pm0.235$. The uncertainty of $r$ becomes nearly four
times larger than the previous one (the result in the case with
fixed $n_t$), due to the free tensor spectral index $n_t$.  The
constraint on $n_t$ is: $\overline{n_t}\pm\Delta
n_{t}=-0.022\pm0.240$.

From FIG.\ref{figure7}, we also find the strong correlation
between $n_t$ and $r_p$, which can be easily understood. It is due
to we have chosen the pivot scale $k_{0}=0.05$Mpc$^{-1}$. However,
the quantity $r$ in this scale, $r(k_0)$,  is not the one which is
measured most precisely. We assume that there is a
tensor-to-scalar ratio $r^*(k_t^*)$ (the tensor-to-scalar ratio at
the pivot wavenumber $k_t^*$), which can be measured most
precisely. We expect this quantity $r^*(k_t^*)$ has no correlation
with $n_t$. In this paper, we call $k_t^*$ as the `best pivot
wavenumber'.  Following Eq. (\ref{k1k0}), we can relate
$r^*(k_t^*)$ and $r(k_0)$ by the following formula
 \begin{eqnarray}\label{nt-relation}
r(k_0)=r^*(k_t^*)\left(\frac{k_0}{k_t^*}\right)^{n_t}.\end{eqnarray}
Since in the calculation, we have adopted the input tensor
spectral index $\hat{n}_t=0$, we have
$\hat{r}^*(k_t^*)=\hat{r}(k_0)=0.3$.  However, the uncertainties
of these two quantities ($r^*$ and $r$) are expected to be
different.

\begin{table}
\caption{ The best pivot wavenumber $k_t^*$, the mean values and
the standard deviations of ($r_p$, $n_t$, $r_p^*$). In the
likelihood analysis, we have considered two free parameters ($r$,
$n_t$).}
\begin{center}
\label{2para}
\begin{tabular}{|c|c|c|c|c|c|}
  \hline
    input $\ell_{\max}$ & output parameter& B & CT& CTE & CTEB \\
  \hline
  $100$  & $k^*_t$(Mpc$^{-1}$) &   $1.26\times10^{-3}$ & $3.13\times10^{-3}$ &
$3.43\times10^{-3}$  & $2.25\times10^{-3}$ \\
  $100$  & $\overline{r^*_{p}}\pm\Delta r^*_p$ &   $0.296\pm0.072$ &
$0.288\pm0.081$ & $0.293\pm0.073$  & $0.300\pm0.049$ \\
  $100$  & $\overline{n_{t}}\pm\Delta n_t$ &   $-0.022\pm0.240$ & $0.051\pm0.475$
& $0.023\pm0.369$  & $-0.003\pm0.166$ \\
  $100$  & $\overline{r_{p}}\pm\Delta r_p$ &   $0.351\pm0.235$ & $0.775\pm1.329$ &
$0.502\pm0.678$  & $0.335\pm0.167$ \\
  \hline
  $500$  & $k^*_t$(Mpc$^{-1}$) &   $1.52\times10^{-3}$ & $2.97\times10^{-3}$ &
$3.75\times10^{-3}$  & $2.37\times10^{-3}$ \\
  $500$  & $\overline{r^*_{p}}\pm\Delta r^*_p$ & $ 0.291\pm0.066$ &
$0.291\pm0.083$ & $0.293\pm0.067$  & $0.298\pm0.046$ \\
  $500$  & $\overline{n_{t}}\pm\Delta n_t$ &  $-0.016\pm0.226$ & $-0.024\pm0.433$
& $-0.019\pm0.327$  & $-0.009\pm0.159$ \\
  $500$  & $\overline{r_{p}}\pm\Delta r_p$ &   $0.352\pm0.249$ & $0.559\pm0.950$ &
$0.399\pm0.414$  & $0.325\pm0.172$ \\
  \hline
\end{tabular}
\end{center}
\end{table}

We use the following steps to search for the best pivot wavenumber
$k_t^*$:

{\bf Step 1} Randomly choose a pivot wavenumber $k'$, which is
different from $k_0$.

{\bf Step 2} Calculate the value of $r_p(k')$ by the formula in
Eq. (\ref{k1k0}).

{\bf Step 3} Project the maxima of the likelihood functions for
$300$ realization into $n_t-r_p(k')$ plane.

{\bf Step 4} In $n_t-r_p(k')$ plane, if $r_p(k')$ correlates with
$n_t$, we iterate the same steps from Step 1. Otherwise, if
$r_p(k')$ has the weakest correlation with $n_t$, we get the
result: $k'=k_t^*$, and $r_p(k')=r_p^*(k_t^*)$.

By these four steps, we find, in ``{\it B}" case, the best pivot
wavenumber is $k_t^*=1.26\times10^{-3}$Mpc$^{-1}$.  In
FIG.\ref{figure7}, we plot the distribution of $n_t-r_p^*$ (left
lower panel). As expected, the correlation between $n_t$ and
$r_p^*$ disappears. We also calculate the average value and
standard deviation of $r_p^*$, which is $\overline{r_p^*}\pm\Delta
r_p^*=0.296\pm0.072$. The average value of $r_p^*$ is equal to the
input value $\hat{r^*}=0.3$ within $1\%$ and hence there is no
evidence for bias. The standard deviation of $r_p^*$ ($\Delta
r_p^*=0.072$) is much smaller than that of $r_p$ ($\Delta
r_p=0.235$), but close to the result $\Delta r_p=0.067$ gotten in
Section \ref{s3}, where only free parameter $r$ is considered.
Hence we conclude that, if we choose the best pivot wavenumber,
the free parameter $n_t$ cannot influence the constraint on the
tensor-to-scalar ratio.

We can also consider the constraints on $r$ and $n_t$ in the other
cases. The distributions of $r_p$ and $n_t$ in the $300$
realization are all plotted in FIG.6 (upper panels). The strong
correlations exist in all these panels. By the exactly same steps,
we can find the best wavenumber $k_t^*$, which are all listed in
Table \ref{2para}.  For example, in ``{\it CT}" case $k_t^*=3.13\times
10^{-3}$Mpc$^{-1}$ and in ``{\it CTE}" case $k_t^*=3.43\times
10^{-3}$Mpc$^{-1}$. In these two cases, the best pivot wavenumbers
are close to each other, which are all much larger than that in
``{\it B}" case. In ``{\it CTEB}" case,  the best pivot wavenumber is
$k_t^*=2.25\times10^{-3}$Mpc$^{-1}$, which is between that in ``{\it B}"
case and that in ``{\it CTE}" case. In all these three cases, the values
of $\Delta r_p^*$ are all very close to that of $\Delta r_p$
gotten in Section \ref{s3}, when only free parameter $r$ is
considered. Hence, we obtain the same conclusion, if we adopt the
best pivot wavenumber, the free parameter $n_t$ cannot expand the
constraint on the tensor-to-scalar ratio (We should mention that,
in the latter work \cite{our3}, we have completely proved this
conclusion, and given the analytic formulae for the best pivot
wavenumber $k_t^*$ and the uncertainty $\Delta n_t$).

From Table \ref{2para}, we also find that, the uncertainty of
$n_t$ is always very large. For example in ``{\it CTEB}" case, the
constraint is $\Delta n_t=0.166$, which is fairly loose for the
determination of the physical model of the early universe.

We have also considered another condition, where a broader range
($\ell_{\max}=500$) simulated data $D_{\ell}^{XX'}$ are used for
the likelihood analysis. The results are all listed in Table
\ref{2para}. As expected, we find that in this condition, the
values of $k_t^*$ $\Delta r_p^*$, $\Delta n_t$ are all close to
those in the previous condition, where only simulated data in
large scale ($\ell_{\max}=100$) are considered.


\subsection{Effect of the free parameters $A_s$, $n_s$\label{s5.2}}

In this subsection,  we shall extend the discussion in Section
\ref{s5.1} to the more general case, where we consider four free
parameters: ($r$, $n_t$, $A_s$, $n_s$). By the simulated data, we
can investigate the effects of free parameters $A_s$ and $n_s$ on
the constraint of $r$. The steps are exactly same with that in the
Section \ref{s5.1}. For the simplification, we shall use the
parameter $A'_{s}$, defined by $A'_{s}\equiv
A_{s}/2.3\times10^{-9}$, instead of $A_s$.

We notice that, the power spectra $C_{\ell}^{BB}$ only depends on
$P_{t}(k)$, but not on  $P_{s}(k)$. Since $P_{t}(k)$ is determined
by the parameters $A_t=(rA_{s})$ and $n_t$, in ``{\it B}" case we cannot
constrain the separate parameters: $r$, $n_t$, $A'_{s}$ and $n_s$.
So in this subsection, we shall not discuss the ``{\it B}" case.

\begin{figure}[t]
\centerline{\includegraphics[width=16cm,height=8cm]{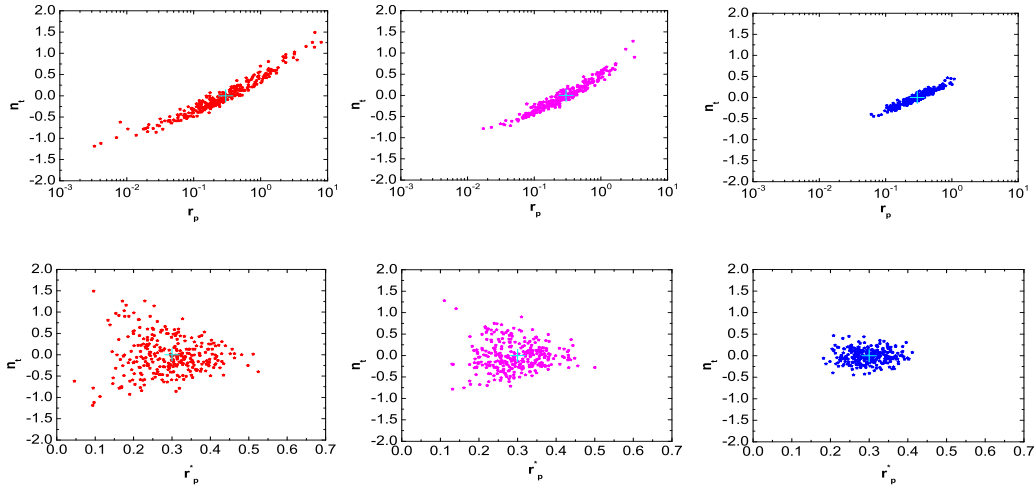}}
\caption{The locations of the maxima from 300 realization
projected into $n_t-r_p$ (upper panels), and $n_t-r^*_p$ (lower
panels) planes. Red (magenta, blue) dots denote the results in
``{\it CT}" (``{\it CTE}", ``{\it CTEB}") case. In all these graphs, we have
considered four free parameters ($r$, $n_t$, $A'_{s}$, $n_s$) in
the likelihood analysis. The input simulated data are up to
$\ell_{\max}=500$, and the sign $``+"$ denotes the input values of
the parameters. }\label{figure8}
\end{figure}

We firstly consider the condition, where the values of
$\ell_{\max}$, $N$, and the input values of parameters are adopted
as in (\ref{coefficients2}). We adopt the pivot wavenumber as in
(\ref{005}). The likelihood functions peak at $(r_p, n_t, A'_{s},
n_s)$. FIG.\ref{figure8} presents the maxima projected into
$n_t-r_p$ plane from $300$ realization. We find the strong
correlation between $n_t$ and $r_p$ exists, which is because we
have used the pivot wavenumber $k_0=0.05$Mpc$^{-1}$. The outputs
$\overline{r_p}\pm\Delta r_p$, $\overline{n_t}\pm\Delta n_t$,
$\overline{A'_{s}}\pm\Delta A'_{s}$, $\overline{n_s}\pm\Delta n_s$
in ``{\it CT}", ``{\it CTE}", ``{\it CTEB}" cases are all listed in Table
\ref{4para}. We find, due to the uncertainties of $A'_{s}$ and
$n_s$, the values of $\Delta r_p$ and $\Delta n_t$ are all larger
than the corresponding results in Table \ref{2para}.

Similar to Section \ref{s5.1}, we can discuss $r^*(k_t^*)$, the
tensor-to-scalar ratio at the best pivot wavenumber $k_t^*$.
Following Eq. (\ref{k1k0}), we can relate $r^*(k_t^*)$ and $r(k_0)$
by the following formula
\begin{eqnarray}\label{nt-relation2}
r(k_0)=r^*(k_t^*)\left(\frac{k_0}{k_t^*}\right)^{n_t-n_s+1}.\end{eqnarray}
Since in the calculation, we have adopted the input spectral index
$\hat{n}_t=0$ and $\hat{n}_s=1$,  we find $\hat{r}^*=\hat{r}=0.3$.
However, the uncertainties of $r^*$ and $r$ are expected to be
different.

We search for the best pivot wavenumber $k_t^*$ by the exactly
same steps, listed in Section \ref{s5.1}. In ``{\it CT}" case, the best
pivot wavenumber is $k_t^*=3.11\times10^{-3}$Mpc$^{-1}$. Based on
this pivot wavenumber, we find $\Delta r^*_p=0.127$. Comparing
with the result of $\Delta r^*_p=0.081$, where only two free
parameters $r$ and $n_t$ are considered, the value of $\Delta
r^*_p$ increases by $60\%$, due to the free parameters $A'_{s}$
and $n_s$. Since we have only used the
simulated in the large scale $\ell\leq 100$ in the likelihood analysis,  the uncertainties of
$A'_s$ and $n_s$ are fairly large (see Table \ref{4para}). This
makes the value $\Delta r^*_p$ is obviously increased.

We have also considered the condition, where $\ell_{\max}=500$ is
adopted. We find the constraints on $A'_{s}$ and $n_s$ become much
smaller: $\Delta A'_{s}=0.005$ and $\Delta n_s=0.008$, and the
constraint on $r_p^*$ becomes $\Delta r^*_p=0.091$, i.e. the
influence of $A'_s$ and $n_s$ on the constraint of $r_p^*$ becomes
much smaller (increasing the value of $\Delta r_p^*$ only by
$10\%$).

\begin{table}
\caption{The best pivot wavenumber $k_t^*$, the mean values and
the standard deviations of ($r_p$, $n_t$, $r_p^*$, $A'_s$, $n_s$).
In the likelihood analysis, we have considered four free
parameters ($r$, $n_t$, $A'_s$, $n_s$). }
\begin{center}
\label{4para}
\begin{tabular}{|c|c|c|c|c|}
  \hline
    input $\ell_{\max}$ & output parameter& CT& CTE & CTEB \\
  \hline
  $100$  & $k^*_t$(Mpc$^{-1}$) & $3.11\times10^{-3}$ & $3.66\times10^{-3}$  &
$2.33\times10^{-3}$ \\
  $100$  & $\overline{r^*_{p}}\pm\Delta r^*_p$ &  $0.291\pm0.127$ &
$0.295\pm0.115$  & $0.300\pm0.069$ \\
  $100$  & $\overline{n_{t}}\pm\Delta n_t$ & $-0.003\pm0.509$ & $0.009\pm0.425$  &
$-0.010\pm0.176$ \\
  $100$  & $\overline{r_{p}}\pm\Delta r_p$ &  $0.783\pm1.693$ & $0.546\pm0.838$  &
$0.331\pm0.178$ \\
  $100$  & $\overline{n_{s}}\pm\Delta n_s$ &  $1.006\pm0.052$ & $1.004\pm0.050$  &
$1.001\pm0.043$ \\
  $100$  & $\overline{A'_{s}}\pm\Delta A'_{s}$ &  $1.016\pm0.092$ &
$1.012\pm0.090$  & $1.006\pm0.078$ \\
  \hline
  $500$  & $k^*_t$(Mpc$^{-1}$) &    $3.02\times10^{-3}$ & $3.35\times10^{-3}$  &
$2.22\times10^{-3}$ \\
  $500$  & $\overline{r^*_{p}}\pm\Delta r^*_p$ &   $0.290\pm0.091$ &
$0.294\pm0.078$  & $0.297\pm0.051$ \\
  $500$  & $\overline{n_{t}}\pm\Delta n_t$ &   $-0.024\pm0.442$ & $-0.053\pm0.291$
 & $-0.016\pm0.155$ \\
  $500$  & $\overline{r_{p}}\pm\Delta r_p$ &   $0.570\pm0.990$ & $0.341\pm0.341$
& $0.316\pm0.166$ \\
  $500$  & $\overline{n_{s}}\pm\Delta n_s$ &  $1.000\pm0.008$ & $0.999\pm0.008$  &
$0.999\pm0.007$ \\
  $500$  & $\overline{A'_{s}}\pm\Delta A'_{s}$ &  $0.999\pm0.005$ &
$0.999\pm0.005$  & $0.999\pm0.005$ \\
  \hline
\end{tabular}
\end{center}
\end{table}

In ``{\it CTE}" and ``{\it CTEB}" cases, we have also investigated the effects
of free parameters $A'_s$ and $n_s$ on the constraint of $r_p^*$.
The results are all similar with those in ``{\it CT}" case. The
$n_t-r_p$ and $n_t-r_p^*$ planes are all plotted in
FIG.\ref{figure8}. We find in both cases, $r_p$ strongly
correlates with $n_t$. However, as expected, $r_p^*$ does not
correlate with $n_t$. The best pivot wavenumber $k_t^*$ and the
constraints of the parameters are all listed in Table \ref{4para}.
Based on these, we conclude that: In the likelihood analysis, if
we only consider the simulated data in the large scale
($\ell\le100$), the constraints of $r^*_p$ and $n_t$ become much
looser, due to the uncertainty of $n_s$ and $A'_{s}$. However, if
we considered the simulated data in the larger range
($\ell\le500$), the constraints on $r^*_p$ and $n_t$ only increase
by $\sim10\%$. Expectable, in the likelihood analysis, if the
simulated data in the range $\ell<2000$ (the real $TT$, $TE$ and
$EE$ data, especially the $TT$ data, in this range are expected to
be well observed by Planck satellite \cite{Planck}) are used,  the
influence of $A'_s$ and $n_s$ on the values of $\Delta r_p^*$ and
$\Delta n_t$ will become negligible.


\section{Conclusion}

The upcoming observations of Planck satellite provide a very
possible opportunity to detect RGWs in the CMB power spectra. In
this paper, by both the simulation and the analytic approximation
methods, we have discussed the detection abilities for RGWs in
four (``{\it B}", ``{\it CT}", ``{\it CTE}", ``{\it CTEB}") cases. The main conclusion can
be summarized as: 1) In ``{\it B}" (``{\it CT}", ``{\it CTE}", ``{\it CTEB}") case, the
Planck satellite can detect the signal of RGWs at 2$\sigma$ level
when $r>0.06$ ($r>0.16$, $r>0.13$, $r>0.05$). 2) Comparing ``{\it CTE}"
with ``{\it B}", we find that, when $r>0.3$, the value of the
signal-to-noise ratio $S/N$ is larger in ``{\it CTE}" case, and when
$r<0.3$, the value of $S/N$ is larger in ``{\it B}" case. If the
realistic noise power spectra of Planck satellite is enlarged for
some reasons, the value of $S/N$ in ``{\it B}" case will be much
reduced. However, in ``{\it CTE}" case, the value of $S/N$ is little
influenced. 3) The value of $S/N$ is much larger in ``{\it CTEB}" case
than that in ``{\it B}" case, especially when $r>0.1$. 4) The free
parameters $n_t$, $n_s$ and $A_s$, cannot reduce the value of
$S/N$, if we consider the data in a large range and adopt the best
pivot scale.

~

~


{\bf Acknowledgement:}

The author thanks D.Baskaran for helpful discussion, and
L.P.Grishchuk for the comment and helpful suggestion on the draft.
This work is partly supported by Chinese NSF under grant Nos.
10703005 and 10775119. In this paper, we have used the CAMB code
\cite{camb} to calculate the CMB power spectra.



\appendix

\section{Gaussian approximations of the likelihood functions}
\label{appendix}

In this appendix, by using the Gaussian approximation of the pdfs
for the estimators $D_{\ell}^{XX'}$, we shall simplify the exact
likelihood functions, given in Section \ref{s2}.

\subsection{Approximation of $\mathcal{L}_{\rm B}$ }

First, let us focus on the analytic approximation of
$\mathcal{L}_{\rm B}$. We use the following Gaussian function
$f_{G}(D_{\ell}^{BB})$ to approximate the exact pdf
$f(D_{\ell}^{BB})$, \begin{eqnarray}
f_{G}(D_{\ell}^{BB}(\hat{r}))=\frac{1}{\sqrt{2\pi} \Delta
D_{\ell}^{BB}(\hat{r})}\exp\left[-\frac{(D_{\ell}^{BB}(\hat{r})-C_{\ell}^{BB})^2}{
2(\Delta D_{\ell}^{BB}(\hat{r}))^2}\right].\end{eqnarray}
Inserting this formula into Eq. (\ref{blikelihood1}), we obtain
that \cite{wishart2}
\begin{eqnarray} \label{blikelihood-ana} \mathcal{L}_{\rm
B}(r)=C\prod_{\ell=2}^{\ell_{\max}}\left\{\frac{1}{\sqrt{2\pi}
\Delta
D_{\ell}^{BB}(\hat{r})}\exp\left[-\frac{(D_{\ell}^{BB}(\hat{r})-C_{\ell}^{BB})^2}{
2(\Delta D_{\ell}^{BB}(\hat{r}))^2}\right]\right\} ,\end{eqnarray}
where $C$ is the constant for the normalization of the likelihood
function, $D_{\ell}^{BB}(\hat{r})$ is the data, based on the input
tensor-to-scalar ratio $\hat{r}$. $\Delta D_{\ell}^{BB}(\hat{r})$
is the standard deviation of $D_{\ell}^{BB}(\hat{r})$. We should
mention that, as a kind of approximation, Eq.
(\ref{blikelihood-ana}) can give results consistent with the exact
likelihood function (the detailed discussion can be found in
\cite{wishart2}).

Up to a constant, we can rewritten the likelihood
(\ref{blikelihood-ana}) as follows:
\begin{eqnarray} \label{blikelihood-ana2} -2\ln\mathcal{L}_{\rm
B}(r)=\sum_{\ell=2}^{\ell_{\max}}\left[\frac{C_{\ell}^{BB}-D_{\ell}^{BB}(\hat{r})}
{\Delta D_{\ell}^{BB}(\hat{r})}\right]^2,\end{eqnarray}  which
includes the variable  $r$ only by the power spectrum
$C_{\ell}^{BB}$.

\subsection{Approximation of $\mathcal{L}_{\rm CTE}$} Before
proceeding on the ``{\it CT}" case, let us firstly focus on the analytic
approximation in ``{\it CTE}" case.  The likelihood function
$\mathcal{L}_{\rm CTE}$ depends on the pdf
$f(D_{\ell}^{TE},f_{\ell}^{TT},C_{\ell}^{EE})$, which is the
Wishart function in Eq. (\ref{wishart}). Similar to the
approximation of $f(D_{\ell}^{BB})$, here we shall use the
following multivariate normal function to approximate the exact
Wishart distribution function:
\begin{eqnarray} \label{f_{G}3}f_{G}(\vec{D_{\ell}})
=\frac{1}{(2\pi)^{3/2}|\Sigma|^{1/2}}
\exp\left[-\frac{1}{2}(\vec{D_{\ell}}-\vec{C_{\ell}})^{\rm
T}\Sigma^{-1}(\vec{D_{\ell}}-\vec{C_{\ell}})\right],
\end{eqnarray}
where the vectors $\vec{D_{\ell}}$ and $\vec{C_{\ell}}$ are
defined as $\vec{D_{\ell}}\equiv[D_{\ell}^{TE}(\hat{r}),
D_{\ell}^{TT}(\hat{r}), D_{\ell}^{EE}(\hat{r})]^{\rm T}$,
$\vec{C_{\ell}}\equiv[C_{\ell}^{TE}, C_{\ell}^{TT},
C_{\ell}^{EE}]^{\rm T} $. $\Sigma$ is the covariance matrix of the
variable $\vec{D_{\ell}}$.  Based on the Gaussian assumption of the
CMB field, the estimators $D_{\ell}^{XX'}$ have covariance as
below (see for instant \cite{covariance,wishart2})
\begin{subequations}
\begin{eqnarray}\label{c1}
 {\rm
cov}(D_{\ell}^{TT},D_{\ell}^{TT})&=&\frac{2(C_{\ell}^{TT}+N_{\ell}^{TT}W_{\ell}^{-
2})^2}{(2\ell+1)f_{\rm sky}},~~\\\ \label{c2}
 {\rm
cov}(D_{\ell}^{EE},D_{\ell}^{EE})&=&\frac{2(C_{\ell}^{EE}+N_{\ell}^{EE}W_{\ell}^{-
2})^2}{(2\ell+1)f_{\rm sky}},~~\\ \label{c3}
 {\rm
cov}(D_{\ell}^{TE},D_{\ell}^{TE})&=&\frac{(C_{\ell}^{TE})^2+(C_{\ell}^{TT}+N_{\ell
}^{TT}W_{\ell}^{-2})(
C_{\ell}^{EE}+N_{\ell}^{EE}W_{\ell}^{-2})}{(2\ell+1)f_{\rm sky}},
~\\ \label{c4}
 {\rm
cov}(D_{\ell}^{TT},D_{\ell}^{EE})&=&\frac{2(C_{\ell}^{TE})^2}{(2\ell+1)f_{\rm
sky}},~~\\ \label{c5}
 {\rm cov}(D_{\ell}^{TE},D_{\ell}^{TT})&=&\frac{2C_{\ell}^{TE}(
C_{\ell}^{TT}+N_{\ell}^{TT}W_{\ell}^{-2})}{(2\ell+1)f_{\rm
sky}},\\ \label{c6}
 {\rm cov}(D_{\ell}^{TE},D_{\ell}^{EE})&=&\frac{2C_{\ell}^{TE}(
C_{\ell}^{EE}+N_{\ell}^{EE}W_{\ell}^{-2})}{(2\ell+1)f_{\rm sky}}.
\end{eqnarray}
\end{subequations}

In order to investigate the cross relation between the estimators,
we define the cross-correlation coefficient as
 \begin{eqnarray}\label{d2}
 \rho_{XX'YY'}\equiv\frac{{\rm cov}(D_{\ell}^{XX'},D_{\ell}^{YY'})}{\sqrt{{\rm
cov}(D_{\ell}^{XX'},D_{\ell}^{XX'}) {\rm
cov}(D_{\ell}^{YY'},D_{\ell}^{YY'})}}.
 \end{eqnarray}
From the relations in Eqs. (\ref{c1}-\ref{c6}), we can obtain that
 \begin{eqnarray}\label{d3}
 \rho_{TTEE}=\rho_{\ell}^2,~~
 \rho_{TETT} =\rho_{TEEE}=
 \rho_{\ell}\sqrt{\frac{2}{1+\rho^2_{\ell}}}~,
\end{eqnarray}
where $\rho_{\ell}$ is expressed in (\ref{rho}), which have been
detailed discussed in our previous paper \cite{ours}.  Taking into
account the Planck instrumental noises, in the large scale
($\ell\leq100$), we have $\rho_{\ell}<0.45$ \cite{ours}. This
makes that the correlation coefficients $\rho_{TTEE}$,
$\rho_{TETT}$ and $\rho_{TEEE}$ are all much smaller than 1. So in
the analytic approximation, we ignore the correlation between
different estimators. Based on this approximation, we can simplify
the multivariate normal function $f_{G}(\vec{D})$ in
(\ref{f_{G}3}) as the following form:
\begin{eqnarray}
\label{f_{G}32}f_{G}(\vec{D_{\ell}})=\prod_{XX'}f_{G}(D_{\ell}^{XX'}(\hat{r})),\end{eqnarray}
where $XX'=TE, TT, EE$. The function
$f_{G}(D_{\ell}^{XX'}(\hat{r}))$ is the following Gaussian
function
\begin{eqnarray} f_{G}(D_{\ell}^{XX'}(\hat{r}))=\frac{1}{\sqrt{2\pi} \Delta
D_{\ell}^{XX'}(\hat{r})}\exp\left[-\frac{(D_{\ell}^{XX'}(\hat{r})-C_{\ell}^{XX'})^
2}{2(\Delta D_{\ell}^{XX'}(\hat{r}))^2}\right].\end{eqnarray}

Inserting the approximation pdf (\ref{f_{G}32}) into Eq.
(\ref{ctelikelihood1}) and ignoring the independent constant, we
get the approximation likelihood function,
\begin{eqnarray} \label{ctelikelihood-ana2} -2\ln\mathcal{L}_{\rm
CTE}(r)=\sum_{\ell=2}^{\ell_{\max}}\sum_{XX'}\left[\frac{C_{\ell}^{XX'}-D_{\ell}^{
XX'}(\hat{r})}{\Delta
D_{\ell}^{XX'}(\hat{r})}\right]^2,\end{eqnarray} where
$XX'=TE,TT,EE$.

\subsection{Approximation of $\mathcal{L}_{\rm CT}$} Let us turn
our attention to the analytic approximation of $\mathcal{L}_{\rm
CT}$. Similar to the discussion of $\mathcal{L}_{\rm CTE}$, we can
get the approximation form of $\mathcal{L}_{\rm CT}$. Up to a
constant, the likelihood is written as
\begin{eqnarray} \label{ctlikelihood-ana2} -2\ln\mathcal{L}_{\rm
CT}(r)=\sum_{\ell=2}^{\ell_{\max}}\sum_{XX'}\left[\frac{C_{\ell}^{XX'}-D_{\ell}^{X
X'}(\hat{r})}{\Delta
D_{\ell}^{XX'}(\hat{r})}\right]^2,\end{eqnarray} where
$XX'=TE,TT$.

\subsection{Approximation of $\mathcal{L}_{\rm CTEB}$}

We can also discuss the approximation form of likelihood
$\mathcal{L}_{\rm CTEB}$. Since $\mathcal{L}_{\rm
CTEB}=\mathcal{L}_{\rm CTE}\mathcal{L}_{\rm B}$, using Eqs.
(\ref{blikelihood-ana2}) and (\ref{ctelikelihood-ana2}), we obtain
that
\begin{eqnarray} \label{cteblikelihood-ana2} -2\ln\mathcal{L}_{\rm
CTEB}(r)=\sum_{\ell=2}^{\ell_{\max}}\sum_{XX'}\left[\frac{C_{\ell}^{XX'}-D_{\ell}^
{XX'}(\hat{r})}{\Delta
D_{\ell}^{XX'}(\hat{r})}\right]^2,\end{eqnarray} where
$XX'=TE,TT,EE,BB$.



\end{document}